\documentclass[lettersize,journal]{IEEEtran}
\usepackage{amsmath,amsfonts, amsthm}

\newtheorem{assumption}{Assumption}

\usepackage{array}
\usepackage[caption=false,font=normalsize,labelfont=sf,textfont=sf]{subfig}
\usepackage{textcomp}
\usepackage{stfloats}
\usepackage{url}
\usepackage{verbatim}
\usepackage{graphicx}
\usepackage{cite}
\usepackage[colorlinks=true, linkcolor=blue, citecolor=blue, urlcolor=blue]{hyperref}
\usepackage[linesnumbered,ruled,vlined]{algorithm2e}
\SetKwInput{KwInput}{Input}
\SetKwInput{KwOutput}{Output}
\usepackage{balance}
\usepackage{color, xcolor}
\usepackage{multicol}
\usepackage{ragged2e}
\usepackage{blkarray}
\usepackage{todonotes}
\usepackage{orcidlink}
\usepackage{booktabs} 
\usepackage[table]{xcolor}
\definecolor{lightgreen}{RGB}{220, 255, 220}

\begin{document}

\title{DSFL: A Dual-Server Byzantine-Resilient Federated Learning Framework via Group-Based Secure Aggregation}
\author{Charuka Herath\,\orcidlink{0009-0009-1369-0120},~\IEEEmembership{Student Member,~IEEE,}
Yogachandran Rahulamathavan,\orcidlink{0000-0002-1722-8621},~\IEEEmembership{Member, ~IEEE,}
Varuna De Silva,\orcidlink{0000-0001-7535-141X},~\IEEEmembership{Member, ~IEEE,}
and Sangarapillai Lambotharan,\orcidlink{0000-0001-5255-7036},~\IEEEmembership{Senior Member, ~IEEE,}

\thanks{ 
C. Herath, Y. Rahulamathavan, Varuna De Silva, and S. Lambotharan are with the Institute for Digital Technologies, Loughborough University, London, U.K. (e-mails: \{c.herath, y.rahulamathavan, v.d.de-silva, s.lambotharan\}@lboro.ac.uk).

Copyright (c) 2024 IEEE. Personal use of this material is permitted. However, permission to use this material for any other purposes must be obtained from the IEEE by sending a request to pubs-permissions@ieee.org.


This work is supported by Engineering and Physical Sciences Research Council (EPSRC) projects under grant EP/X012301/1, EP/X04047X/2, and EP/Y037243/1

}

}




\maketitle

\begin{abstract}
Federated Learning (FL) enables decentralised model training without sharing raw data, offering an attractive privacy-preserving paradigm. However, current FL protocols struggle to defend against Byzantine participants, preserve model utility under non-independent and identically distributed (non-IID) data, and remain computationally feasible on edge devices. Prior approaches either assume trusted hardware, rely on heavy cryptographic tools, utilise extreme hardware capabilities, or fail to address simultaneous threats to robustness and privacy. In this work, we propose DSFL, A Dual-Server Byzantine-Resilient FL Framework via Group-Based Secure Aggregation that overcomes these limitations through a novel, non-cryptographic design. Unlike LSFL, which assumes non-colluding, semi-honest servers, our framework eliminates this dependency by exposing and addressing a key vulnerability: the possibility of privacy leakage through collusion between clients and servers. DSFL introduces three core innovations: (1) a dual-server secure aggregation protocol that preserves update confidentiality without encryption or pairwise key exchange, (2) a group-wise, credit-based filtering mechanism to identify and suppress Byzantine participants based on deviation scoring, and (3) a dynamic reward–penalty strategy to enforce fair, adaptive participation over training rounds. DSFL is evaluated on MNIST, CIFAR-10, and CIFAR-100 under up to 30\% Byzantine participants and across independent and identically distributed (IID) and non-IID settings. It consistently outperforms state-of-the-art baselines, including LSFL, HE-based protocols, and DP-enhanced schemes. For example, DSFL achieves 97.15\% accuracy on CIFAR-10 and 68.60\% on CIFAR-100, while FedAvg degrades to 9.39\% under moderate adversarial conditions. Despite its strong resilience, DSFL remains lightweight, introducing only 55.9 ms runtime and 1088 KB communication overhead per round, making it suitable for secure, scalable deployment in edge computing environments.
\end{abstract}


\section{Introduction}
\label{Section: Introduction}

\IEEEPARstart{T}{he} rapid proliferation of intelligent edge devices ranging from LLM-enabled assistants to IoT sensors and autonomous participants has led to a growing demand for privacy-preserving collaborative intelligence. While centralised machine learning systems continue to dominate, recent regulatory attention from bodies like the Global Privacy Assembly \cite{GPA2023} has underscored the risks of training large models on centralised datasets, which are often noisy, incomplete, or privacy-invasive \cite{DataIssues2023, cai2015challenges}.  Furthermore, implementing controls to address the data protection risks posed by the use of these datasets is very challenging. Moreover, \cite{GPA2023} states that if not properly secured, ML model outputs can reveal sensitive or private information included in the datasets used for training, leading to potential or real data breaches.

Federated Learning (FL) \cite{McMahan2016, JakubMcMahan2016} has emerged as a compelling solution to this challenge. FL enables distributed participants to jointly train a global model without revealing raw data, offering enhanced privacy and regulatory compliance. Unlike traditional ML, where data is centralised \cite{TradML2021, centralizeFL2020}, \cite{centralizeFL2021}, FL participants and the server collaboratively train a model. Real-world FL deployments span use cases from keyword recognition in voice assistants \cite{McMahan2016} to cross-silo model validation in medical AI \cite{karargyris2023medperf} and autonomous vehicle collaboration \cite{nvidia2024}. However, despite its promise, FL remains vulnerable to critical risks that undermine its viability in participantic workflows.

We group these critical risks into three categories. Firstly, the privacy leakage of local model updates remains a primary concern, where adversaries exploit shared gradients to infer sensitive participant data. These leakages can intermediate gradients that could unintentionally reveal information about the training data. This can occur without access to raw data and reveals a core vulnerability in the FL model \cite{multiRoundPrivacy2023, Melis2018}, \cite{ge2023review, FLsecAgg2022}. Byzantine adversaries in FL participants that send corrupted updates can cause model compromise, compromising both model integrity and data privacy. For instance, model replacement attacks \cite{modelreplacing2025, backdoor}, \cite{replacement2019} allow an attacker to completely override the global model with malicious intent through a single update and expose the vulnerability of standard aggregation schemes like FedAvg that fail to bound or normalise updates. Backdoor attacks \cite{ dualBackdoor, backdoorICRL} have shown even more subtle manipulation: Participants can cause the model to misclassify specific inputs while maintaining overall accuracy. Label-flipping attacks, where adversarial clients deliberately mislabel data to mislead the learning process, are shown by \cite{modePoisioning2020} to be particularly destructive in non-independent and identically distributed (non-IID) settings. These attacks are subtle yet potent, as they can bypass typical validation due to the poisoned data’s structural similarity to legitimate samples. Moreover, GAN-based attacks represent a severe breach of privacy, as illustrated by \cite{gan2024, Hitaj2017}, \cite{gan2025}, where adversaries can exploit shared gradients to reconstruct private training data using generative adversarial networks, revealing a critical link between robustness and privacy in FL. Moreover, these attacks are often successful even under secure communication, as the leakage stems from the model update itself. This has alarming implications for applications in healthcare and finance, where data exposure can violate legal regulations like HIPAA and GDPR. 

Secondly, FL assumes that all participants behave honestly and upload well-trained local models. In practice, this assumption breaks down due to poisoning attacks \cite{fang2020local, backdoor}, \cite{replacement2019}. A small fraction of malicious clients can significantly degrade the accuracy and reliability of the global model \cite{tolpegin2020data, jagielski2018manipulating}. Various defence mechanisms have been proposed \cite{blanchard2017machine, guerraoui2018hidden}, \cite{mean_median, munoz2019byzantinerobust}, including Byzantine-robust aggregation and anomaly detection. However, many of these mechanisms rely on computationally intensive non-linear operations, making them incompatible with privacy-preserving secure computation.

Finally, almost all the current secure aggregation protocols are communication-intensive and require elaborate key exchanges, making them unsuitable for edge devices. For instance, Google's Secure Aggregation (SecAgg) \cite{aggregation2017} requires each client to establish pairwise keys with all other clients and perform multiple rounds of cryptographic operations, resulting in communication and computational overheads that scale quadratically with the number of clients. Similarly, homomorphic encryption-based protocols like Paillier and CKKS \cite{He2018, He2017} support computation on encrypted data but suffer from heavy ciphertext expansion and arithmetic complexity. Consequently, achieving a practical, privacy-preserving, and Byzantine-robust FL framework remains an open challenge, especially for resource-constrained edge environments \cite{wahab2021federated}.

\subsection{Motivation}
These challenges underscore a fundamental tension in practical FL deployments: how to simultaneously achieve privacy preservation, Byzantine robustness, and computational efficiency, particularly on edge devices with limited resources. Existing methods often prioritise one goal at the expense of others. For instance, approaches using homomorphic encryption or differential privacy offer strong privacy guarantees but introduce prohibitive overheads and degrade model utility in non-IID or adversarial settings. Conversely, lightweight schemes may lack provable privacy or robustness, making them unsuitable for adversarial environments. This persistent trade-off necessitates the design of FL protocols that are secure, robust, and practical for real-world deployment.

As a solution, et al. introduced a promising direction in the form of a dual-server secure aggregation protocol called LSFL \cite{LSFL}. This approach aims to provide Byzantine resilience while avoiding heavy encryption-based computations by leveraging additive noise and non-colluding semi-honest servers. While conceptually efficient, LSFL suffers from a critical security limitation: its privacy protection can be breached if a malicious client colludes with one of the aggregation servers—an assumption that is difficult to uphold in untrusted environments. As we detail in Section~\ref{Section: System Model and Problem Formulation}, the protocol is vulnerable to a simple reconstruction attack that exposes individual model updates using shared aggregate information.

Motivated by this limitation, we propose an improved design that builds on LSFL’s architecture but addresses its core vulnerabilities through formal enhancements. Specifically, our approach DSFL, introduces a secure, modular secret-sharing scheme and a trust-aware, group-based aggregation mechanism. These additions reduce collusion risk and strengthen both privacy and robustness under adversarial conditions while maintaining low computational and communication overhead, making it particularly suited for edge-based FL deployments.

As shown in our evaluations, DSFL outperforms existing schemes across multiple dimensions—privacy, Byzantine tolerance, and scalability—without sacrificing convergence speed or accuracy. Table~\ref{Table: Key feature comparison} summarises these benefits, and a detailed technical comparison against relevant baselines is presented.

\begin{table*}[t]
\centering
\caption{Key features and challenges addressed by DSFL compared to prior FL frameworks.}
\label{Table: Key feature comparison}
\scriptsize
\begin{tabular}{@{}l l l c c c c c c c c@{}}
\toprule
\textbf{Algorithm} & \textbf{Reference} & 
\begin{tabular}[c]{@{}c@{}}High Accuracy\\ (Byzantine)\end{tabular} & 
\begin{tabular}[c]{@{}c@{}}Fast\\ Convergence\end{tabular} & 
\begin{tabular}[c]{@{}c@{}}Lightweight\\ (No Encryption)\end{tabular} & 
\begin{tabular}[c]{@{}c@{}}Fairness\\ Guarantee\end{tabular} & 
\begin{tabular}[c]{@{}c@{}}Low Comm. \&\\ Comp. Cost\end{tabular} &
\begin{tabular}[c]{@{}c@{}}Malicious \\ Privacy\end{tabular} &
\begin{tabular}[c]{@{}c@{}}Poisoning \\ Resilience\end{tabular} &
\begin{tabular}[c]{@{}c@{}}Server\end{tabular}\\
\midrule
FedAvg               & \cite{McMahan2016}     &  & \checkmark & \checkmark &  & \checkmark &  &  & single\\
SecAgg               & \cite{aggregation2017} &  & \checkmark & \checkmark &  & \checkmark & \checkmark & & single\\
Trimmed Mean         & \cite{mean_median}     &  & \checkmark & \checkmark &  & \checkmark & \checkmark & & single\\
Median               & \cite{mean_median}     &  & \checkmark & \checkmark &  & \checkmark & \checkmark & & single\\
FLOD                 & \cite{FLOD}            & \checkmark & \checkmark &  &  & \checkmark & \checkmark & \checkmark & dual\\
DPBFL                & \cite{DPBFL}           & \checkmark &  &  &  & \checkmark & \checkmark  & \checkmark & single\\
HE-based FL          & \cite{He2018}          & \checkmark &  &  & \checkmark &  & \checkmark & \checkmark & single\\
ELSA                 & \cite{elsa}            & \checkmark & \checkmark &  &  & \checkmark & \checkmark & \checkmark & dual\\
LSFL                 & \cite{LSFL}            & \checkmark & \checkmark & \checkmark &  & \checkmark & \checkmark & & dual\\
\rowcolor{lightgreen}
\textbf{DSFL (Ours)} & This paper            & \checkmark & \checkmark & \checkmark & \checkmark & \checkmark &  & \checkmark & sual\\
\bottomrule
\end{tabular}
\end{table*}

\subsection{Contributions and Paper Organisation}

Initial work \cite{LSFL} motivated this work and introduced a novel grouping mechanism that enhances LSFL’s security without sacrificing efficiency. DSFL is designed to support collaborative model training among edge nodes while ensuring low overhead, resistance to adversarial attacks, and strong data privacy guarantees. Our simulation results show that the proposed approach, DSFL, achieves SOTA performance in terms of security, privacy, and computational efficiency. Moreover, Table \ref{Table: Key feature comparison} states the challenges in existing work in brief, and a detailed comparison of the challenges can be found in Table \ref{Table: Performance Comparison}.

The key contributions of this paper are as follows:

\begin{enumerate}
    \item We introduce DSFL, a privacy-preserving, FL framework that improves edge service quality without compromising user data.
    \item We improved the originally proposed LSFL by \cite{LSFL}, proposing a Dual-Server Byzantine-Resilient FL Framework via Group-Based Secure Aggregation protocol by introducing a novel (participant combination matrix) PCM and a contributed participant group matrix (CPG). It can achieve secure aggregation and Byzantine robustness by addressing the security vulnerability. Expressly, it preserves the privacy information of the participants in the Byzantine robustness and model aggregation phases.
    \item We implemented a prototype of DSFL and evaluated its performance in terms of training accuracy, training time, and Byzantine robustness. Our results show that DSFL can achieve the fidelity, efficiency and security design goals.
    \item We conduct rigorous experiments showcasing DSFL’s robustness against both data and model poisoning attacks. Furthermore, we provide a convergence analysis, an Efficiency and Overhead Analysis, and a security Analysis.
\end{enumerate}
\subsection{Paper Organization}

The subsequent sections of this paper are organised as follows. Section \ref{Related Work} summarises the related work in the same research area to address the three main problems. Section \ref{Section: System Model and Problem Formulation} states the system model and problem scope and provides why traditional dual server architecture (LSFL) \cite{LSFL} have been compromised. Section \ref{Section: Proposed Methodology} explains the proposed DSFL Framework. Section \ref{Section: Experiments and Evaluation} states the results and discussion on the proposed approach, which improves robustness against byzantine participants, the convergence speed and model accuracy, provides an efficiency and overhead analysis, a convergence guarantee and a security guarantee. Section \ref{Section: Conclusion} serves as the conclusion of the paper, summarising the key findings, contributions and future work. Additionally, all notations are listed in Table \ref{Table: Table of Notations}.


\section{Related Work}
\label{Related Work}

The development of FL has prompted extensive investigation into secure aggregation, Byzantine robustness, and communication efficiency. We categorise related works into three subtopics: (i) FL with single aggregators, (ii) FL with distributed trust, and (iii) other defence strategies.

\subsection{FL with Single Aggregators}

Many FL frameworks adopt a single aggregator architecture where the central server collects local model updates from participants. This setup, while straightforward, introduces significant risks related to single points of failure and privacy leakage. To address Byzantine attacks in such centralised settings, aggregation techniques such as Krum \cite{KRUM}, Trimmed Mean, and Median \cite{mean_median} have been proposed. These statistical methods filter outliers or select updates based on distance metrics but require direct access to raw gradients, undermining privacy.

Secure aggregation protocols aim to mitigate such leakage. Bonawitz et al. proposed SecAgg \cite{aggregation2017}, where clients share encrypted model updates using pairwise masks. However, this incurs quadratic communication overhead and complex key negotiation. Homomorphic encryption (HE)-based methods such as Aono et al. \cite{He2018} and BatchCrypt \cite{BatchCrypt} offer stronger confidentiality but are computationally expensive for edge devices. VerifyNet \cite{Xu2019} enhances SecAgg with verifiability, while SEAR \cite{zhao2022sear} leverages Trusted Execution Environments (TEEs) for stronger integrity, though TEEs introduce hardware dependencies and scalability concerns.

\subsection{FL with Distributed Trust}

To overcome the central aggregator’s limitations, recent efforts have shifted toward decentralised trust models. These include dual-server setups or fully decentralised FL. The LSFL framework \cite{LSFL} introduces a lightweight dual-server aggregation scheme that improves scalability and robustness under semi-honest assumptions but remains vulnerable to collusion. Building upon LSFL, ELSA \cite{elsa} proposes input validation with range proofs and cross-checking between two non-colluding servers, enabling enhanced privacy and robustness. ELSA also evaluates dropout resilience and incorporates defences against malicious clients without requiring direct access to updates. Zhao et al. \cite{zhao2022pvdfl} propose a verifiable decentralised FL system (PVD-FL) using zero-knowledge proofs to validate updates and aggregation correctness. Although privacy-preserving, these verification layers introduce substantial computational costs. Likewise, SEFL \cite{sun2022sefl} incorporates federated blockchains and multi-server shuffling to prevent gradient leakage but requires additional trust coordination. The rest of the works \cite{prio, secByz2020} only provide privacy against a semi-honest server, and therefore, leave much to be desired, but still, LSFL colludes, and ELSA is less efficient.

\subsection{Other Defense Methods}

Beyond architectural improvements, statistical and cryptographic defences have been developed to address poisoning, backdoor, and gradient inference attacks. Liu et al. \cite{liu2021privacy} and Zhou et al. \cite{zhou2023dpfl} utilise differential privacy (DP) to perturb gradients, thus hiding sensitive information. However, DP often degrades model utility due to noise. Mu~{n}oz-Gonz'alez et al. \cite{munoz2019byzantinerobust} introduce adaptive model averaging techniques that improve resilience to data poisoning. Moreover, Byzantine-resilient model selection frameworks have also been explored. The MKrum algorithm \cite{blanchard2017machine} and its variants attempt to estimate model trustworthiness using distance metrics. More integrated solutions have emerged. Dong et al. \cite{dong2024p2brofl} proposed the $\Pi_{P2Brofl}$ framework, which uses a 3-party secure computation model to simultaneously ensure privacy and robustness. Their top-k protocol enables efficient top-k model selection under privacy constraints, reducing the cost of secure model filtering. Their results indicate minimal model degradation even under 50\% Byzantine presence. However, such MPC-based systems may still be too resource-heavy for edge environments. Other approaches attempt to combine robustness and efficiency. FedMD \cite{FedMD2019} applies model distillation for FL with non-IID data, enhancing generalisation and reducing communication. However, it requires public auxiliary datasets and lacks robust aggregation under adversarial scenarios.

Despite significant progress, many prior works either rely on strong cryptographic tools, incur heavy computation and communication costs, or require unrealistic trust assumptions. Our DSFL framework is designed to be both practical and secure. It leverages dual-server lightweight aggregation, integrates secret sharing for privacy, and filters poisoned clients through group-wise credit estimation—making it suitable for scalable FL on edge devices.

\begin{table}[t]
    \flushright
    \centering
    \scriptsize
    \caption{Table of Notations}
    \centering
    \begin{tabular}{cc|cc}
    \toprule
    $P$ & Local Participant & $F(\cdot)$ &  Local Objective Functions\\
    \midrule
    $\mathbf{w}$ & Local Model & $\epsilon $& Gaussian Noise\\
    $N$ & Total Local Devices & $\bar{\textbf{w}}$ & Global Model\\
    $d$ & Euclidean distance & $E$ & Local epochs\\
    $b$ & Local mini-batch size & $L(\cdot)$ & Loss Function\\
    $k$ & threshold & $\eta$ & Learning Rate \\
    $\gamma$ & Credit & $\phi$ & Penalty\\
    $\varsigma$ & Reward & $\sigma$ & Cost\\
    $m$ & $PCM_{i,j}$ rows & $P^*$ & Malicious Participants\\
    $G_m$ & $PCM_{i,j}$ groups & $M$ & median\\
    $\mathbf{P}$ & Selected Participants & $\mathcal{O}()$ & Big O \\
    \bottomrule
    \end{tabular}
    \label{Table: Table of Notations}
\end{table}
\section{System Model and Problem Formulation}
\label{Section: System Model and Problem Formulation}

\subsection{System Entities and Assumptions}
The LSFL system comprises three main entities, as depicted in Figure~\ref{fig:systemOverview}, the Third-Party Service Provider (TP), the Service Provider (SP), and the Local participants ($P$s).

The TP is a third-party service provider responsible for selecting eligible participants for each training round and initiating the PCM. It filters out suspicious or potentially Byzantine nodes based on predefined criteria to ensure that only reliable participants contribute to model aggregation. By doing so, it enhances the integrity of the aggregation process handled by the SP.

The SP acts as the edge computing service provider and maintains the initial global model. It seeks to improve the model using the decentralised data and computational resources of the participating edge devices. The SP coordinates the training process, initialises CPG aggregates local model updates, and maintains the latest global model.

The $P$s are local participants, such as mobile phones or IoT devices, that collect private data through on-device sensors. These devices contribute to the FL process by training local models and uploading updates to the SP. However, $P$s are not assumed to be fully trustworthy. Some may behave incorrectly or maliciously, contributing inappropriate or adversarial updates to the training process.

We adopt an honest-but-curious model for both the SP and TP. While these entities follow the protocol correctly and do not actively disrupt the training process, they may attempt to infer private information from the updates received from $P$s. This assumption reflects realistic commercial motivations, where service providers and third-party authorities may seek to extract sensitive insights from user data, even without direct malicious intent.

\subsection{Byzantine Device Agents}
In practical FL deployments, edge devices are susceptible to adversarial behaviour due to limited oversight and computational constraints. Following prior work~\cite{blanchard2017machine, elsa}, we assume that most participants are honest, but a minority may behave maliciously. Specifically, we consider two types of Byzantine $P$s:
(1) Free-riders: These are passive participants who avoid computation by submitting arbitrary or noisy updates, often simulated using Gaussian noise~\cite{AccurateDib2018}. While not overtly malicious, they degrade model quality without directly attacking it.
(2) Malicious nodes: These adversaries aim to poison the global model by submitting crafted updates. We model this using label-flipping attacks~\cite{dualBackdoor}, where the adversary inverts the direction of the update to counteract convergence.

\subsection{Problem Formulation}
This section revisits the LSFL scheme proposed in \cite{LSFL} and analyses a security vulnerability that exposes user model updates to an adversarial server.
In \cite{LSFL}, user $i$ partitions the local model update $\mathbf{w}_i$ into two components, $\mathbf{w}_i^{(1)}$ and $\mathbf{w}_i^{(2)}$, as follows:

\begin{equation}
\mathbf{w}_i^{(1)} = \mathbf{w}_i + \epsilon_i, \quad \mathbf{w}_i^{(2)} = \mathbf{w}_i - \epsilon_i,
\end{equation}

Where $\epsilon_i$ is Gaussian noise introduced to obscure the partitioning. The original model update can be reconstructed as:

\begin{equation}
\mathbf{w}_i = \frac{1}{2}\mathbf{w}_i^{(1)} + \frac{1}{2}\mathbf{w}_i^{(2)}.
\end{equation}

The primary objective of \cite{LSFL} is to ensure that model updates remain concealed from the participating servers. In this scheme, user $i$ transmits $\mathbf{w}_i^{(1)}$ to the TP and $\mathbf{w}_i^{(2)}$ to the SP. Upon collecting updates from all $N$ users, TP and SP compute:

\begin{equation}
z_1 = \sum_{i=1}^N \mathbf{w}i^{(1)}, \quad z_2 = \sum{i=1}^N \mathbf{w}_i^{(2)},
\end{equation}

Where $z_1$ is exclusively known to TP and $z_2$ is known only to SP. Subsequently, TP transmits $z_1$ to SP, enabling SP to derive the global model update:

\begin{equation}
\mathbf{\bar{w}} = \frac{1}{2N}(z_1 + z_2).
\end{equation}

Using $\mathbf{\bar{w}}$, SP calculates the distance metric $d_i^{(2)}$ between $\mathbf{w}_i^{(2)}$ and $\mathbf{\bar{w}}$:

\begin{equation}
\label{eq:distance}
d_i^{(2)} = \frac{1}{2}\mathbf{w}_i^{(2)} - \mathbf{\bar{w}}, \quad \forall i.
\end{equation}

SP then transmits all $d_i^{(2)}$ values to TP. The security analysis in \cite{LSFL} claims that, assuming TP and SP do not collude, TP cannot infer user model updates. However, the following section demonstrates a vulnerability that enables TP to reconstruct user updates by exploiting the transmitted $d_i^{(2)}$ values.

The security of the LSFL scheme \cite{LSFL} is contingent on the assumptions that (i) TP and SP do not engage in collusion and (ii) the majority of users are non-malicious. However, in practical FL environments with numerous users, collusion between a subset of users and a server is a plausible scenario. A privacy-preserving FL scheme must, therefore, be resilient to such adversarial behaviours.

A secure FL scheme should withstand collusion attacks such that a significant number of users must conspire with a server to breach privacy. However, if a single user’s collusion with a server can compromise the entire scheme, as is the case with LSFL \cite{LSFL}, the system is critically vulnerable.

To illustrate this attack, consider a scenario where the first user is adversarial and shares its partitioned update $\mathbf{w}_1^{(2)}$ with TP. TP can leverage this information along with the received distance values \eqref{eq:distance} from SP to reconstruct the remaining users’ model updates. Specifically, TP computes:

\begin{equation}
\label{eq:calculated_split}
\mathbf{w}_i^{(2)} = 2(d_i^{(2)} - d_1^{(2)}) + \mathbf{w}_1^{(2)}, \quad i = 2, \ldots, N.
\end{equation}

The correctness of this reconstruction follows from:

\begin{equation}
d_i^{(2)} - d_1^{(2)} = \frac{1}{2}\mathbf{w}_i^{(2)} - \frac{1}{2}\mathbf{w}_1^{(2)}, \quad i = 2, \ldots, N.
\end{equation}

With $\mathbf{w}_i^{(1)}$ already received from users, TP can now compute the original model update for each user:

\begin{equation}
\mathbf{w}_i = \mathbf{w}_i^{(1)} + \mathbf{w}_i^{(2)}, \quad i = 2, \ldots, N.
\end{equation}

This analysis confirms that the LSFL privacy mechanism \cite{LSFL} can be compromised if a single user colludes with one of the servers, highlighting a critical flaw in the scheme’s security model. While we assume non-colluding servers, prior systems, such as Prio/Prio+ \cite{prio}, make the same assumption to achieve low-overhead masking. Our system inherits this practical balance while ensuring computational feasibility for edge settings


\begin{figure*}[!t]
    \centering
    \includegraphics[width=1\linewidth]{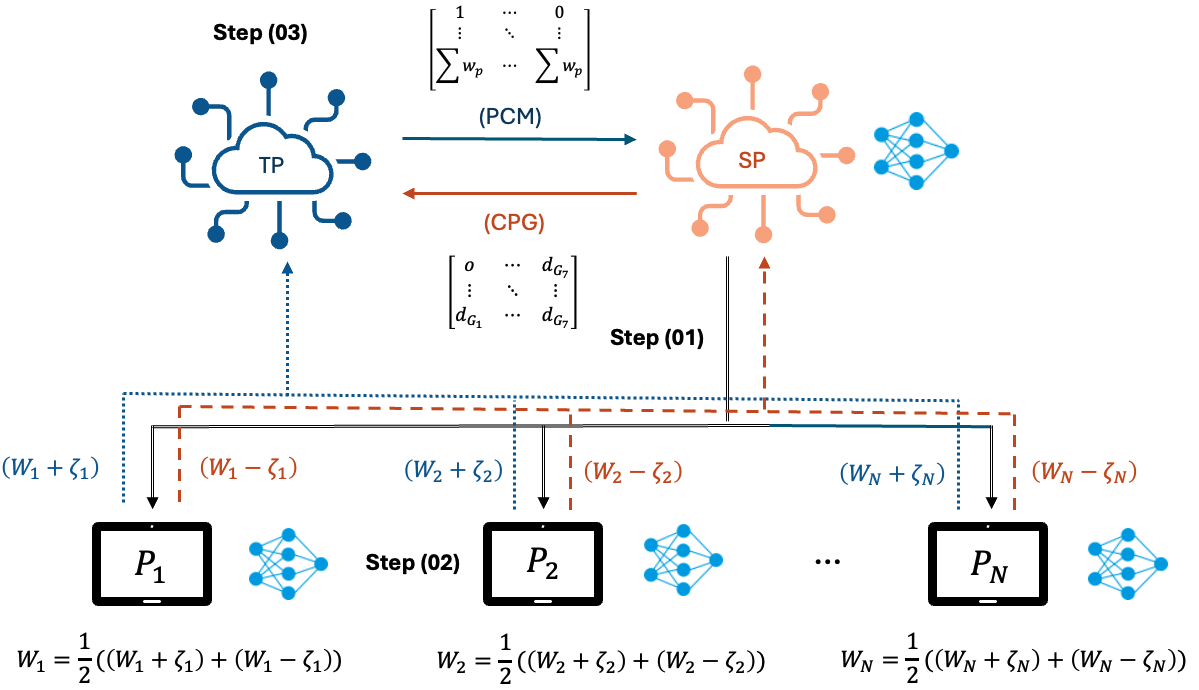}
    \caption{System Overview}
    \label{fig:systemOverview}
\end{figure*}


\section{Proposed DSFL Framework}
\label{Section: Proposed Methodology}


\begin{algorithm}[t]
\caption{Private Local Model Training}
\label{Algorithm: Local Model}
\KwIn{learning rate $\eta$, loss $L(w, D)$, epochs $E$, noise variance $g^2$}
Initialize global model $w_0$ \;
\For{$i = 1$ to $E$}{
    Sample batch $D_i$ from $D$
    $\mathbf{w}_i \leftarrow \mathbf{w}_{i-1} - \eta \nabla L(\mathbf{w}_{i-1}, D_b)$ \\
    Generate random noise $\zeta_i \sim \mathcal{N}(0, g^2)$\;
    Compute: $\mathbf{w}_i^{(1)} = \mathbf{w}_i - \zeta_i$, $\mathbf{w}_i^{(2)} = \mathbf{w}_i + \zeta_i$
}
Send $\mathbf{w}_i^{(1)}$ to TP and $\mathbf{w}_i^{(2)}$ to SP \;
\KwOut{Secret shares $\mathbf{w}_i^{(1)}$, $\mathbf{w}_i^{(2)}$}
\end{algorithm}

A comprehensive outline of the system is depicted in Figure \ref{fig:systemOverview}. In each iteration, $P$s, TP, and SP engage in FL training. To maintain clarity, $P$s are denoted as $P$, represented as a set $\{P_1, P_2, \dots, P_N\}$. The primary objective of each $P$ is to train a machine learning model locally while ensuring the confidentiality of their datasets from external entities within the system.

FL operates cyclically, encompassing the following three phases:
(depicted in Figure \ref{fig:systemOverview}):
\begin{algorithm}[t]
    \label{Algorithm: Aggregation}
    \SetAlgoLined 
    \caption{Dual-Server Byzantine-Resilient FL Framework via Group-Based Secure Aggregation}
    \KwIn{Secret shares $\mathbf{w}_i^{(1)}, \mathbf{w}_i^{(2)}$, credit $\gamma$, penalty $\phi$, reward $\varsigma$, cost $\sigma$, participants $P$, threshold $k$}
    \KwOut{global model parameter $W$}
    Initialize: $\phi$, $\gamma$, $\varsigma$, $\sigma$, $k$\
    $\forall P_i$, TP receive $\textbf{w}^{(1)}_i$ and SP receive $(\mathbf{w}_i)^{(2)}$\
    // Secure Byzantine Robustness\;
    Compute $z_1 = \frac{1}{N} \sum \mathbf{w}_i^{(1)}$, $z_2 = \frac{1}{N} \sum \mathbf{w}_i^{(2)}$
    TP randomly groups participants into $m$ groups\;
    TP computes the $\sum_{p=1}^{P} \mathbf{w}_{p}^{(1)}$ for each group\;
    TP computes $PCM_{i,j}$\;
    TP sends its share of $PCM_{i,j}$ to SP\;
    SP computes $\bar{\mathbf{w}}$\;
    SP computes $d_{G_i} =  \parallel \bar{\mathbf{w}} - \bar{\mathbf{w}}_{G_i} \parallel ^2_2$\ for each $G_i$\; 
    SP computes each $\sum d_{P_i}$\;
    SP set the $CPG_{i,j}$\ and transmit back to TP;
    $d' \in \{\sum d_{P_1}, \sum d_{P_2}, ..., \sum d_{P_N}\}$\;
    TP gets the median $M$ of $d'$\;
    \For{all $P_i \in P$}{
        TP computes $d''_i = |\sum d_{P_1} - M|$\;
    }
    TP chooses the top $k$ participants with the smallest similarity $P \leftarrow \{P_1, P_2,..., P_P\}$\;
    // Secure Fairness Guarantee\;
    $i \leftarrow 1$\;
    \While{$i \leq N$}{
        \If{$P_i \in P$}{
            $\gamma_i \leftarrow \gamma_i + \varsigma - \sigma$\ ;
        }
        \Else{
            $\gamma_i \leftarrow \gamma_i - \phi - \sigma$\;
        }
        \If{$\gamma_i < 0$}{
            $P.\text{remove}(P_i)$\;
        }
        $i \leftarrow i + 1$\;
    }
    // Secure Model Aggregation\;
    TP computes $\mathbf{w}^{(1)} = \frac{1}{2k} \sum_{P_i \in \mathbf{P}} \mathbf{w}^{(1)}_i$\;
    TP sends the new subset of groups $\{G_1\}$ in the PMC\;
    SP retrieves the corresponding masked shares $\{\mathbf{w}^{(2)}_i \mid P_i \in \mathbf{P}\}$\;
    SP computes $\mathbf{w}^{(2)} = \frac{1}{2k} \sum_{P_i \in \mathbf{P}} \mathbf{w}^{(2)}_i$\;
    SP computes $W \leftarrow \textbf{w}^{(1)} + \textbf{w}^{(2)}$\;
    SP reveals $W$ to participants $P$, and updates the global model\;
\end{algorithm}

\textbf{Step I: Global Model Synchronisation with Selected Participants}
The SP maintains the initial model to facilitate collaborative training among FL participants. At the beginning of each training round, SP transmits the most recent global model $\mathbf{w}_{\text{global}}$ to the selected participants $P \in N$. Each participant stores this model locally for subsequent training iterations.

\textbf{Step II: Local Model Training}
Upon receiving the global model, each selected participant trains a local model by refining the global model using its respective local dataset. Each participant $P_i$ possesses a local dataset $D_i$, where $i \in \{1, 2, \dots, N\}$, such that $D = \sum_{i=1}^{N} D_i$. The aggregators, TP and SP, retain portions of the global model parameters. The optimisation problem for each participant $P_i$ is defined as:
\begin{equation}
\min_{\mathbf{w}_i} L(D_i, \mathbf{w}_i),
\end{equation}
where $L(D_i, \mathbf{w}_i)$ represents the empirical loss on training dataset $D_i$, and $\mathbf{w}_i$ denotes participant $P_i$’s local model parameters.


Unlike conventional FL schemes, model parameter computations are distributed among participants to (i) reduce the computational burden on the aggregator and (ii) leverage the computing resources of participants. To enhance model parameter security and mitigate security risks associated with aggregators, a dual-server architecture is adopted for secure aggregation, similar to the approach proposed in \cite{LSFL}. Each participant $P_i$ splits its private parameter $\mathbf{w}_i$ into two additive secret shares:
\begin{equation}
\mathbf{w}_i = \frac{1}{2} (\mathbf{w}_i^{(1)} + \mathbf{w}_i^{(2)}) = \frac{1}{2} (\mathbf{w}_i + \zeta_i + \mathbf{w}_i - \zeta_i),
\end{equation}
where $\zeta_i$ is Gaussian noise sampled from $\mathcal{N}(0, \sigma^2)$ with $\sigma = 20$. Each participant $P_i$ sends $\mathbf{w}_i^{(1)}$ and $\mathbf{w}_i^{(2)}$ to SP and TP, respectively. The dual-Server secure model update process follows this step. This process is systematically delineated in Algorithm \ref{Algorithm: Local Model}.

\textbf{Step III: Dual-Server Secure Model Update:} 
The reconstruction of the correct global model by the SP is contingent upon the collaborative efforts of the TP and SP. The SP is responsible not only for preserving the model parameters but also for overseeing secure aggregation. 


To ensure the security of model updates and the preservation of data privacy, SP and TP collaboratively compute the global model $W$ through secure aggregation of secret-shared local model updates, conforming to a predefined aggregation rule. This rule incorporates Secure Byzantine Robustness, Secure Fairness Guarantee, and Secure Model Aggregation principles, ensuring that SP and TP (i) remain oblivious to the update parameters of any individual participant, (ii) mitigate Byzantine attacks that could compromise the accuracy of the global model, and (iii) encourage participant engagement while maintaining fairness among them.

Upon receiving the secret shared model parameters from each participant $P_i$, the aggregators initiate the process of updating the global model and subsequently disseminate the updated model parameters to all participants. 



As described in Algorithm \ref{Algorithm: Aggregation}, the proposed aggregation protocol comprises three integral components: secure Byzantine robustness, secure fairness guarantee, and secure model aggregation. Importantly, neither TP nor SP can reconstruct the individual model updates $\mathbf{w}_i$ from the aggregated group results, due to the linear under-determined nature of the problem.

\subsubsection{\textbf{Secure Byzantine Robustness}}

To counteract the influence of Byzantine participants on the global model, it is crucial to scrutinise the model parameters submitted by each participant and select the most reliable ones. This study proposes a k-nearest weight selection approach that effectively resists Byzantine participants.

Initially, TP and SP independently compute the summations $\sum_{i=1}^{N} \mathbf{w}_{i}^{(1)}$ and $\sum_{i=1}^{N} \mathbf{w}_{i}^{(2)}$, respectively. Subsequently, TP constructs a Participant Combination Matrix ($\text{PCM}_{ij}$) of dimensions $N \times G$, where $G_i$ represents the $i^{th}$ group among a total of $m$ groups. Such that: 
\[
\text{PCM}_{ij} = \begin{cases} 1 & \text{if participant } P_i \in G_j \\ 0 & \text{otherwise} \end{cases}
\]
The value of $m$ is determined based on the percentage of malicious participants ($P^*$) in each training round, $m$ is computed as:
$
    m = \left| \frac{N - \frac{P^*}{P} \times 100}{10} \right| - 1.
$
If $P^*$ is set to $20$, then $\text{PCM}_{ij}$ looks like below.
\[
\begin{blockarray}{c@{\hspace{1em}}ccccccc}
    & \text{G}_1 & \text{G}_2 & \text{G}_3 & \text{G}_4 & \text{G}_5 & \text{G}_6 & \text{G}_7 \\
\begin{block}{c[ccccccc]}
    \text{P}_1 & 1 & 1 & 1 & 0 & 0 & 1 & 0 \\
    \text{P}_2 & 1 & 0 & 1 & 0 & 0 & 0 & 1 \\
    \text{P}_3 & 1 & 0 & 0 & 0 & 1 & 0 & 0 \\
    \text{P}_4 & 1 & 0 & 1 & 0 & 0 & 0 & 0 \\
    \text{P}_5 & 1 & 0 & 0 & 1 & 0 & 0 & 0 \\
    \text{P}_6 & 1 & 1 & 0 & 0 & 0 & 1 & 1 \\
    \text{P}_7 & 1 & 0 & 0 & 0 & 1 & 0 & 1 \\
    \text{P}_8 & 1 & 1 & 0 & 1 & 0 & 0 & 0 \\
    \text{P}_9 & 1 & 0 & 0 & 0 & 0 & 1 & 0 \\
    \text{P}_{10} & 1 & 0 & 0 & 1 & 1 & 0 & 0 \\
\end{block}
    & \sum_{p=1}^{P} \mathbf{w}_{p}^{(1)} & \sum & \sum & \sum & \sum & \sum & 
    \sum
\end{blockarray}
\]

For the baseline experimental setting, we set $N = 10$ participants and $P^* = 20$ (assuming 20\% of participants are malicious), and the parameter $m$ is set to 7. Each group $G_i$ consists of three user combinations of selected participants, with $G_1$ serving as a special case that includes all participants. Finally, TP calculates $\sum_{p=1}^{P} \mathbf{w}_{p}^{(1)}$ for each $G$ group, generating the finalized output PCM. When the $N$ for each training round increases, the $m$ and PCM dimensions change accordingly.

Crucially, when $m < N$ or $\text{rank}(\text{PCM}) < N$, this system is underdetermined, i.e., there are more unknowns than equations and hence has infinitely many solutions. Even when $m \geq N$, the deliberate overlap and redundancy in group definitions prevent a unique inversion of $\mathbf{w}^{(1)}$ or $\mathbf{w}^{(2)}$ without access to both shares. Thus, TP (or SP) alone, having access only to $\mathbf{w}^{(1)}$ (or $\mathbf{w}^{(2)}$), and the PCM, cannot uniquely solve for the original $\mathbf{w}_i$ vectors.

This design ensures information-theoretic privacy under a semi-honest, non-colluding assumption between TP and SP. Even if TP or SP attempts linear inference using all available groupwise aggregates and the known PCM structure, the resulting system remains underdetermined and unsolvable without the complementary share. As the number of participants grows, the PCM's sparsity and overlapping structure further obscure the contribution of any single participant, enhancing both privacy and robustness.

Subsequently, the TP transmits the computed sums of local model weights \( \sum_{p=1}^{P} \mathbf{w}_{p}^{(i)} \), where \(i \in \{1,...m\} \) (denoted as \( \bar{\mathbf{w}_{G_i}} \)) for each group \( G_i \), along with the PCM, to the SP.

The SP then computes the average model \( \bar{\mathbf{w}} \) by eliminating the influence of each participant's local model parameters. This is formally expressed as:
\[
\bar{\mathbf{w}} = \frac{1}{2N} \left( \sum_{i=1}^{N} \mathbf{w}_{i}^{(1)} + \sum_{i=1}^{N} \mathbf{w}_{i}^{(2)} \right)
\]
Additionally, the SP determines the Euclidean distance \( d_{G_i} \) between the average model \( \bar{\mathbf{w}} \) and each group-specific average model \( \bar{\mathbf{w}_{G_i}} \):
\[
 d_{G_i} =  \parallel \bar{\mathbf{w}} - \bar{\mathbf{w}}_{G_i} \parallel ^2_2
\]

Given PCM and distances $d_{G_i}$, the Contributed participant Group matrix $\text{CPG}_{ij}$ is defined as:
\[
\text{CPG}_{ij} = \begin{cases} d_{G_i} & \text{if } \text{PCM}_{ij} = 1 \\ 0 & \text{otherwise} \end{cases}
\]
Each row sum $\sum d_{P_i}$ in the CPG provides a score for participant $P_i$ reflecting its contribution’s consistency with the majority.

Following this computation, the cells of the CPG matrix containing ones are updated with their corresponding distances \( d_{G_i} \), leading to the revised CPG matrix representation. 

\[
\vspace{-0.5cm}
\begin{blockarray}{c@{\hspace{1em}}ccccccc}
    & \text{G}_2 & \text{G}_3 & \text{G}_4 & \text{G}_5 & \text{G}_6 & \text{G}_7 & \\
\begin{block}{c[cccccc]c}
    \text{P}_1 & d_{G_2} & d_{G_3} & 0 & 0 & d_{G_6} & 0 & \sum d_{P_1}\\
    \text{P}_2 & 0 & d_{G_3} & 0 & 0 & 0 & d_{G_7} & \sum d_{P_2}\\
    \text{P}_3 & 0 & 0 & 0 & d_{G_5} & 0 & 0 & \sum d_{P_3}\\
    \text{P}_4 & 0 & d_{G_3} & 0 & 0 & 0 & 0 & \sum d_{P_4}\\
    \text{P}_5 & 0 & 0 & d_{G_4} & 0 & 0 & 0 & \sum d_{P_5}\\
    \text{P}_6 & d_{G_2} & 0 & 0 & 0 & d_{G_6} & d_{G_7} & \sum d_{P_6}\\
    \text{P}_7 & 0 & 0 & 0 & d_{G_5} & 0 & d_{G_7} & \sum d_{P_7}\\
    \text{P}_8 & d_{G_2} & 0 & d_{G_4} & 0 & 0 & 0 & \sum d_{P_8}\\
    \text{P}_9 & 0 & 0 & 0 & 0 & d_{G_6} & 0 & \sum d_{P_9}\\
    \text{P}_{10} & 0 & 0 & d_{G_4} & d_{G_5} & 0 & 0 & \sum d_{P_{10}}\\
\end{block}
\end{blockarray}
\]

The updated CPG matrix is then transmitted back to the TP, serving as a criterion for assessing similarity and ensuring the integrity of the model update process while identifying potentially malicious model updates.

The presence of outliers in \( \bar{\mathbf{w}}_{G_i} \), introduced by malicious participants submitting large model update parameters, may lead to deviations from the global model \( \bar{\mathbf{w}} \). Furthermore, an increased \( \sum d_{P_i} \) may indicate the presence of such participants. These outliers can be identified as nodes responsible for deviations in the direction of the global model \( \bar{\mathbf{w}} \).

To quantify deviations, TP computes the set \( d'\), where \( d' \in \{\sum d_{P_1}, ..., \sum d_{P_N}\} \) and derives the median value \( M \) of \( d' \). Subsequently, TP calculates 
\[
d''_i = |\sum d_{P_1} - M|
\]
for each \( P_i \in P \). This calculation reveals the distance between \( \mathbf{w}_i \) and \( M \), establishing a criterion for assessing participant similarity.

Based on these computed deviations, participants are selected for global model aggregation. The parameter \( k \) represents the number of participants whose updates are considered most similar to the collective behaviour of the majority and are thus deemed trustworthy for global model aggregation. In this context, the similarity is quantified using the L2 distance between each participant's update and the median deviation \( M \) computed across all participants. Smaller deviations imply closer alignment with the majority and lower likelihood of Byzantine behaviour. 

Selecting an appropriate value of \( k \) is critical: too small a value may exclude legitimate updates and slow convergence, while too large a value risks including adversarial updates. In our protocol, \( k \) is set empirically based on system parameters such as the expected proportion of malicious participants \( P^* \), the total number of participants \( N \), and empirical sensitivity analysis during preliminary training rounds. Typically, \( k \) is chosen such that it satisfies \( k \leq N - P^* \), ensuring that no more than the expected number of malicious participants are included. For example, with \( N = 10 \) and \( P^* = 4 \) (20\% malicious), we set \( k = 8 \), ensuring that only the 80\% most reliable participants are aggregated.

\subsubsection{\textbf{Collaborative Filtering Method}}

A collaborative filtering approach is integrated to further enhance the robustness of the aggregation process. This method adjusts participants' weights based on their similarity to high-performing participants. Leveraging collaborative filtering techniques, TP assigns higher weights to those whose models have demonstrated strong performance in past iterations.

TP arranges the values of \( d_{p_i} \) in ascending order, such that participant similarity decreases as the values increase. Based on \( d_{p_i} \), the selected participants \( \mathbf{P} \) are determined by selecting the top \( k \)-nearest participants' model updates \( \bar{\mathbf{w}_p^i} \), which are then used to compute the global model. The parameter \( k \) denotes the number of top-ranked participants (based on deviation scoring) selected for aggregation in each round. In practice, \( k \) is chosen such that \( k \leq N - \lfloor \beta N \rfloor \), where \( \beta \in [0, 0.2   ] \) reflects the estimated upper bound on the fraction of Byzantine participants. This ensures that only the most consistent and trustworthy updates contribute to the global model. To maintain robustness while preserving convergence, we empirically set \( k = \lfloor (1 - \beta)N \rfloor \). This choice guarantees the exclusion of potential adversarial updates while still providing sufficient coverage of honest updates for convergence, as theoretically justified in similar robust FL frameworks \cite{yin2018byzantinerobust, blanchard2017machine}.

Inspired by prior work \cite{LSFL}, a system of rewards and penalties is introduced to regulate participant behaviour during model updates. The primary objective of this mechanism is to distinguish between beneficial and detrimental participants, ensuring that those who contribute positively to the global model are incentivised while those who do not are penalised.

To effectively quantify the impact of individual participants on the convergence of the global model, each participant's contribution is systematically evaluated. A participant $P_i$ is classified as beneficial if their updates contribute constructively to the global aggregation process. Conversely, participants whose updates do not positively influence the model convergence are categorised as detrimental. The core principle governing this mechanism is to rigorously assess whether $P_i$ facilitates the convergence of the global model $W$.

To enforce this mechanism, a reward-penalty framework is introduced. If a participant $P_i$ enhances the global model's performance, they are assigned a reward $\varsigma$. Otherwise, they incur a penalty. To ensure fairness and prevent exploitation, an initial allocation of computational credits, denoted as $\gamma$, is uniformly distributed among all participants before the commencement of training. Participants are then required to allocate a fixed cost $\sigma$ per training iteration.

The evaluation of each participant's contribution is derived from their selection during the Byzantine Robustness phase. If a participant is not selected for aggregation, this implies that their updates were deemed unreliable or uninformative, indicating that they have only received the global model without substantively contributing to its refinement. As a consequence, such participants are penalised, leading to a gradual depletion of their allocated credits. Persistent non-contributory behaviour results in fund exhaustion, thereby disqualifying them from further participation in FL. This systematic exclusion process effectively mitigates the influence of Byzantine adversaries on global model aggregation.

The implementation of this framework involves an iterative credit-based validation system. Initially, the aggregation server distributes computational credits uniformly across all participants. During each training round, the server verifies participants' credit balances, disseminates the global model, and deducts $\sigma$ from each participant's credit allocation. Participants execute local training and submit model updates, following which the server evaluates their contributions. Those deemed beneficial receive $\varsigma$ as an incentive, while non-contributory participants incur penalties. Over multiple iterations, this scheme ensures that participants with minimal or adversarial contributions are gradually phased out as their credits deplete, fostering an equitable and robust FL environment.

\subsubsection{\textbf{lightweight Secure Aggregation}}

The proposed lightweight secure aggregation protocol enables privacy-preserving global model updates through a dual-server architecture comprising the TP and the SP. Each participant’s model update $\mathbf{w}_i$ is secret-shared into two components: TP receives $\mathbf{w}^{(1)}_i$ and SP receives $\mathbf{w}^{(2)}_i$, such that $\mathbf{w}_i = \mathbf{w}^{(1)}_i + \mathbf{w}^{(2)}_i$.

After executing the Byzantine-robust participant selection phase, TP identifies the set of top-$k$ reliable participants, denoted as $\mathbf{P}$, based on similarity metrics and reward-credit mechanisms. TP then aggregates the partial updates of the selected participants as:
\[
\mathbf{w}^{(1)} = \frac{1}{2k} \sum_{P_i \in \mathbf{P}} \mathbf{w}^{(1)}_i.
\]
Simultaneously, TP determines the subset of groups $\{G_1\}$ associated with these participants from the PCM, and sends both the participant indices $\mathbf{P}$ and the aggregated group identifier $\sum_{p \in \mathbf{P}} G_1$ to SP.

Upon receiving this information, SP retrieves the corresponding masked shares $\{\mathbf{w}^{(2)}_i \mid P_i \in \mathbf{P}\}$ and computes:
\[
\mathbf{w}^{(2)} = \frac{1}{2k} \sum_{P_i \in \mathbf{P}} \mathbf{w}^{(2)}_i.
\]
SP then cancels out the contribution of previously computed placeholder values $\zeta_i$ (used in intermediate deviation analysis) to finalise the aggregation. The global model $W$ is reconstructed as:
\[
W = \mathbf{w}^{(1)} + \mathbf{w}^{(2)}.
\]

The SP updates the global model using $W$ and broadcasts it to all active participants for the next training round. This architecture ensures that neither TP nor SP can individually reconstruct any participant’s complete model update, thereby guaranteeing information-theoretic privacy. The process remains lightweight and efficient, avoiding cryptographic complexity while preserving robust aggregation and data confidentiality.

\section{Simulation Results}
\label{Section: Experiments and Evaluation}

\begin{table*}[t]
\centering
\caption{Test accuracy (\%) under 20\% Byzantine participants for various aggregation protocols across datasets (mean ± std).}
\label{Table: Performance Comparison}
\begin{tabular}{@{}lcccccc@{}}
\toprule
\begin{tabular}[c]{@{}c@{}}Aggregation\\ Protocol\end{tabular} & 
\begin{tabular}[c]{@{}c@{}}MNIST \\ (IID)\end{tabular} & 
\begin{tabular}[c]{@{}c@{}}MNIST \\ (non-IID)\end{tabular} & 
\begin{tabular}[c]{@{}c@{}}CIFAR-10 \\ (IID)\end{tabular} & 
\begin{tabular}[c]{@{}c@{}}CIFAR-10 \\ (non-IID)\end{tabular} &
\begin{tabular}[c]{@{}c@{}}CIFAR-100 \\ (IID)\end{tabular} & 
\begin{tabular}[c]{@{}c@{}}CIFAR-100 \\ (non-IID)\end{tabular}\\
\midrule
FedAvg~\cite{McMahan2016}      & 80.67 ± 0.09 & 78.75 ± 0.12
& 52.56 ± 0.31 & 55.16 ± 0.25 & 8.85 ± 0.27 & 28.10 ± 0.22 \\
SecAgg~\cite{aggregation2017}  & 80.34 ± 0.11 & 79.85 ± 0.34 & 53.39 ± 0.82 & 56.21 ± 0.45 & 8.98 ± 0.89 & 29.34 ± 0.34 \\
Trimmed Mean~\cite{mean_median}& 80.05 ± 0.31 & 72.65 ± 0.40 & 57.39 ± 0.44 & 60.87 ± 0.19  & 55.70 ± 0.33 & 50.43 ± 0.20  \\
Median~\cite{mean_median}      & 89.87 ± 0.10 & 87.19 ± 0.15 & 70.78 ± 0.28 & 76.58 ± 0.22 & 52.20 ± 0.21 & 56.95 ± 0.25 \\
FLOD~\cite{FLOD}               & 88.12 ± 0.07 & 86.12 ± 0.14 & 71.05 ± 0.23 & 77.05 ± 0.20 & 52.52 ± 0.18 & 57.71 ± 0.19 \\
DPBFL~\cite{DPBFL}             & 90.12 ± 0.07 & 88.12 ± 0.14 & 72.05 ± 0.23 & 78.05 ± 0.20 & 53.52 ± 0.18 & 56.71 ± 0.19 \\
ELSA~\cite{elsa}               & 93.45 ± 0.34 & 90.64 ± 0.09 & 97.01 ± 0.24 & 94.34 ± 0.08 & 67.28 ± 0.31 & 56.91 ± 0.34 \\
LSFL~\cite{LSFL}               & 95.77 ± 0.11 & 90.31 ± 0.11 & 97.53 ± 0.33 & 95.25 ± 0.18 & 69.08 ± 0.20 & 59.60 ± 0.75 \\
HE~\cite{He2018}               & 96.01 ± 0.13 & 91.23 ± 0.08 & 96.12 ± 0.21 & 94.71 ± 0.28 & 68.19 ± 0.43 & 58.17 ± 0.34 \\
\rowcolor{lightgreen}
DSFL (Ours)                   & 96.12 ± 0.11 & 90.12 ± 0.08 & 97.15 ± 0.21 & 95.05 ± 0.18 & 68.60 ± 0.20 & 61.18 ± 0.17 \\
\bottomrule
\end{tabular}
\end{table*}

\begin{table}[t]
\centering
\small
\caption{Test accuracy (\%) of DSFL and FedAvg under varying Byzantine ratios on MNIST, CIFAR-10, and CIFAR-100 (IID / non-IID).}
\label{tab:test_accuracy_against_attackers}
\begin{tabular}{@{}lccc@{}}
\toprule
\begin{tabular}[c]{@{}c@{}}Method\\ (\% Byzantine)\end{tabular} &
\textbf{MNIST} & \textbf{CIFAR-10} & \textbf{CIFAR-100} \\
\midrule
FedAvg (0\%)   & 98.67 / 95.65 & 98.16 / 97.56 & 55.16 / 53.56 \\
FedAvg (10\%)  & 72.05 / 90.65 & 92.39 / 70.87  & 37.87 / 9.39  \\
DSFL (10\%)   & 97.87 / 91.19 & 98.58 / 95.78 & 68.85 / 59.40 \\
\rowcolor{lightgreen}
DSFL (20\%)   & 96.12 / 90.12 & 97.15 / 95.05 & 68.60 / 61.18 \\
DSFL (30\%)   & 94.08 / 90.97 & 97.25 / 92.53 & 68.03 / 60.98 \\
\bottomrule
\end{tabular}
\end{table}

\begin{table*}[!ht]
    \centering
    \caption{Comparison of Computation and Communication Overhead Between Our DSFL, HE \cite{Wei2020}, and FedAvg \cite{McMahan2016}}
    \label{tab:overhead_comparison}
    \begin{tabular}{lccc}
        \toprule
        \textbf{Phase} & \textbf{Our DSFL} & \textbf{HE \cite{Wei2020}} & \textbf{FedAvg \cite{McMahan2016}} \\
        \midrule
        \textbf{Computation} & $\mathcal{O}(T_{tr}) + \mathcal{O}(mT_{ns}) + \mathcal{O}(2mT_{add})$ & $\mathcal{O}(T_{tr}) + \mathcal{O}(mT_{enc})$ & $\mathcal{O}(E \cdot n \cdot w) + \mathcal{O}(m \cdot w)$ \\ 
        \textbf{Communication} & $\mathcal{O}(2m|w|)$ & $\mathcal{O}(m|w_{enc}|)$ & $\mathcal{O}(2m|w|)$ \\ 
        \bottomrule
    \end{tabular}
    \begin{flushleft}
        \footnotesize * $m$ is the number of participants, $w$ is the number of model parameters, and $E$ is the number of local training epochs. $T_{tr}$ is the time overhead for local model training, $T_{ns}$ is the time for noise generation, and $T_{enc}$ is the time for HE. $T_{add}$ represents addition time, and $T_{mul}$ represents multiplication time. $|w|$ and $|w_{enc}|$ denote the size of plain and encrypted weights, respectively.
    \end{flushleft}
\end{table*}

%
%

\subsection{Experimental Setup}

\begin{figure*}[ht]
    \centering
    \includegraphics[width=0.9\linewidth]
    {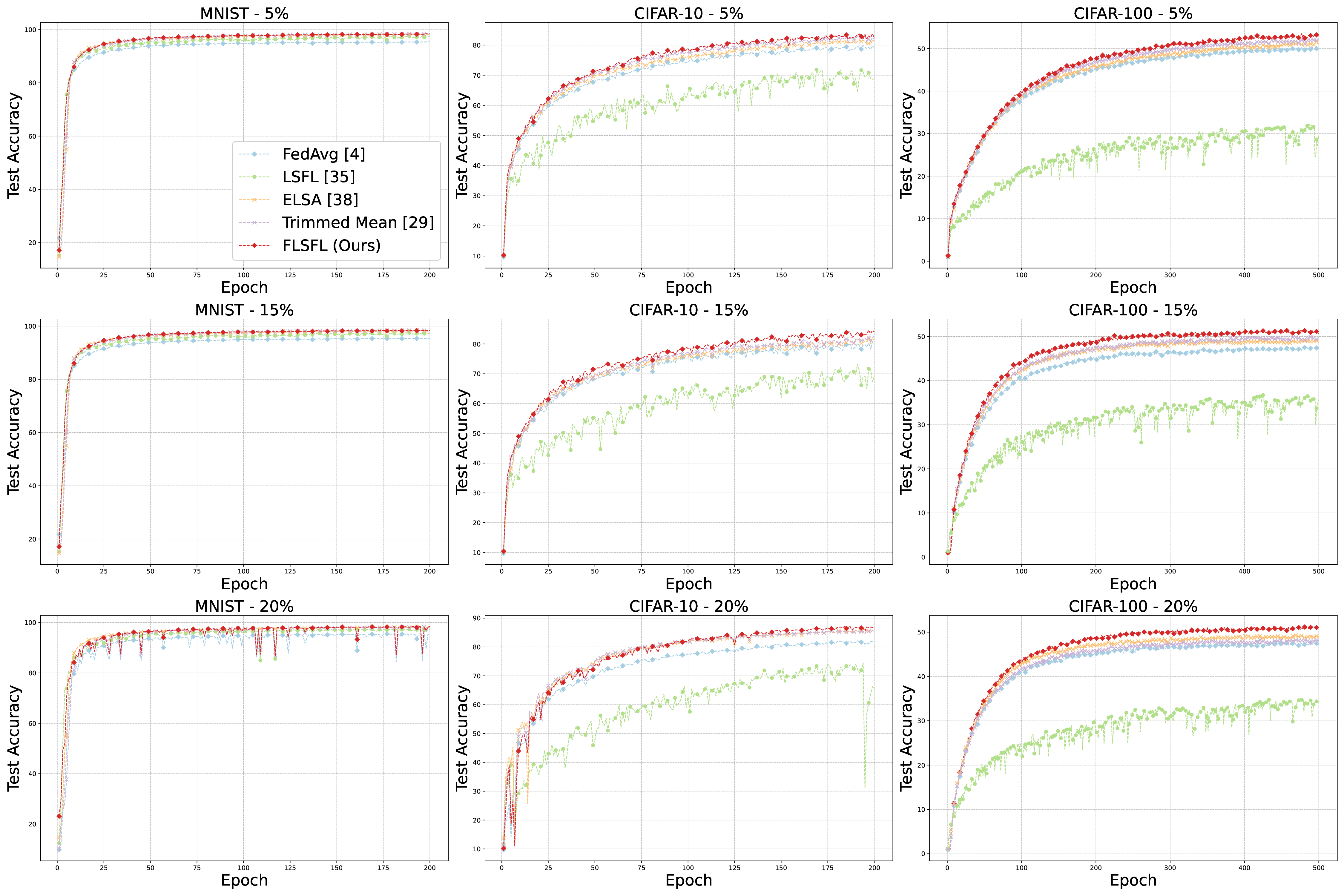}
    \caption{Test accuracies of DSFL and Baseline models for different for 5\%, 10\% and 20\%of Byzantine participants (under label flipping attack) on MNIST, CIFAR-10 and CIFAR-100 datasets (non-IID)}
    \label{fig: DSFL_comparison_all}
\end{figure*}

The experiments were conducted on a high-performance computing setup, utilising an NVIDIA RTX 6000 GPU with 48GB of VRAM, coupled with an Intel Core i9-10980 processor. This hardware configuration ensured efficient training and evaluation of our models, providing reliable and indicative results.

In the experimental setup, we utilised three distinct datasets, MNIST, CIFAR-10, and CIFAR-100 to evaluate the performance of our models. MNIST comprises 60,000 training images and 10,000 test samples, each depicting handwritten grayscale digits ranging from 0 to 9 in a $28\times28$ format. CIFAR-10 consists of 60,000 $32\times32$ colour images distributed across 10 classes. Each class contains 6,000 images, with 5,000 for training and 1,000 for testing. CIFAR-100, on the other hand, contains 60,000 colour images of size $32\times32$ as well, but spread across 100 fine-grained object classes, with 500 training images and 100 testing images per class. 

For MNIST, a Convolutional Neural Network (CNN) was employed, featuring two convolution layers with 32 and 64 kernels of size $3\times3$, followed by a max-pooling layer and two fully connected layers with 9216 and 128 neurons. CIFAR-10 and CIFAR-100 both utilised the ResNet-18 architecture to accommodate the increasing complexity and diversity of classes in these datasets. ResNet-18 was chosen for its ability to effectively handle deeper representations required for learning fine-grained image distinctions, especially in CIFAR-100.

We partitioned each dataset in two ways: IID and non-IID. For the IID setting, the datasets were shuffled and evenly distributed among all participants. For the non-IID setting, the MNIST dataset was divided into 200 shards, each containing 300 examples, with two shards allocated to each participant. Similarly, non-IID versions of CIFAR-10 and CIFAR-100 were constructed by sorting the data by class and allocating a limited subset of classes to each participant, thereby simulating real-world statistical heterogeneity across clients.

\subsection{Adversary Model}

We consider a common but powerful poisoning strategy, label-flipping attacks, where adversarial participants intentionally corrupt their local datasets by mislabeling class outputs. In each federated round, a selected fraction of participants, denoted by $\beta \in \{0.0, 0.1, 0.2, 0.3\}$ (representing 0\%, 10\%, 20\% and 30\% of the total), behave maliciously by flipping labels in their local training data. Specifically, a clean input-label pair $(x, y)$ is altered to $(x, y')$ such that $y' \neq y$, with $y'$ selected to maximise misclassification impact. This induces gradient updates that mislead the global model’s optimisation path.

In addition to label manipulation, malicious clients in our simulation submit poisoned model updates $\mathbf{v}_i = q \cdot \hat{\mathbf{v}}_i$, where $\hat{\mathbf{v}}_i$ is the locally trained update and $q = -1$ inverts the gradient direction to cause maximal model degradation. These adversarial levels are consistent with real-world federated deployments, where compromised edge devices, insider threats, or Sybil attacks can constitute a substantial portion of the network \cite{fang2020local, tolpegin2020data}. For instance, research in adversarial FL suggests even 10\%–20\% Byzantine presence can be sufficient to compromise training \cite{replacement2019}. Our experiments at 20\% and 30\% offer a comprehensive stress test.

To further reflect practical dynamics, the participant pool changes in each round, simulating conditions where clients may intermittently join or drop out due to connectivity or availability, common in edge and IoT scenarios. Malicious participants are randomly sampled from the current pool in each round, ensuring a dynamic and unpredictable adversary presence.

\subsection{Baseline Aggregation Methods}

To comprehensively evaluate the performance of our proposed framework (DSFL), we compared it against several state-of-the-art aggregation protocols under a 20\% Byzantine adversarial setting. These baselines span traditional, statistical, privacy-preserving, and robust aggregation methods.

FedAvg~\cite{McMahan2016} serves as the standard FL baseline without any robustness or privacy safeguards. It performs naive averaging of local models received from clients. SecAgg~\cite{aggregation2017} introduces secure aggregation through cryptographic masking and pairwise key sharing to ensure that the server cannot learn individual model updates. However, it does not provide robustness against adversarial updates. Median, and Trimmed Mean~\cite{mean_median} are statistical aggregation rules aimed at mitigating the impact of outliers and adversaries:
(1) Median selects the element-wise median of local model updates.
(2) Trimmed Mean excludes the highest and lowest $x\%$ of updates and computes the mean of the remaining ones.
FLOD~\cite{FLOD} is a Byzantine-robust approach combining local anomaly detection with aggregation filtering. It detects and removes poisoned model updates before aggregation. DPBFL~\cite{DPBFL} employs Differential Privacy to protect individual model updates while introducing robustness to noisy and adversarial updates. However, its performance often degrades on complex datasets due to the injected noise. HE-based aggregation~\cite{He2018} utilises homomorphic encryption to support secure computation over encrypted data. While offering strong privacy guarantees, it imposes significant computational overhead. ELSA~\cite{elsa} implements a dual-server aggregation scheme with input validation and range proofs to prevent malicious contributions. It enhances both robustness and privacy but assumes partial trust in server separation. LSFL~\cite{LSFL} proposes a lightweight aggregation protocol using two non-colluding servers and random group selection to reduce communication overhead and improve scalability. DSFL (Ours) improves upon LSFL by incorporating a group-based participant filtering mechanism, trust-weighted aggregation, and a reward-penalty fairness mechanism. It achieves superior performance across all datasets, particularly in non-IID and CIFAR-100 settings.

For all experiments, we adopted consistent local training hyperparameters across baselines: a learning rate of 0.001, SGD momentum of 0.5, batch size of 10, and 5 local training epochs per round. The Byzantine participants (either 2 or 5 depending on the scenario) performed label-flipping attacks by submitting malicious updates of the form \( v_i = q \hat{v}_i \), where \( q \) is a negative scalar.

\subsection{Experimental Results}

\begin{figure*}
    \centering
    \includegraphics[width=1\linewidth]{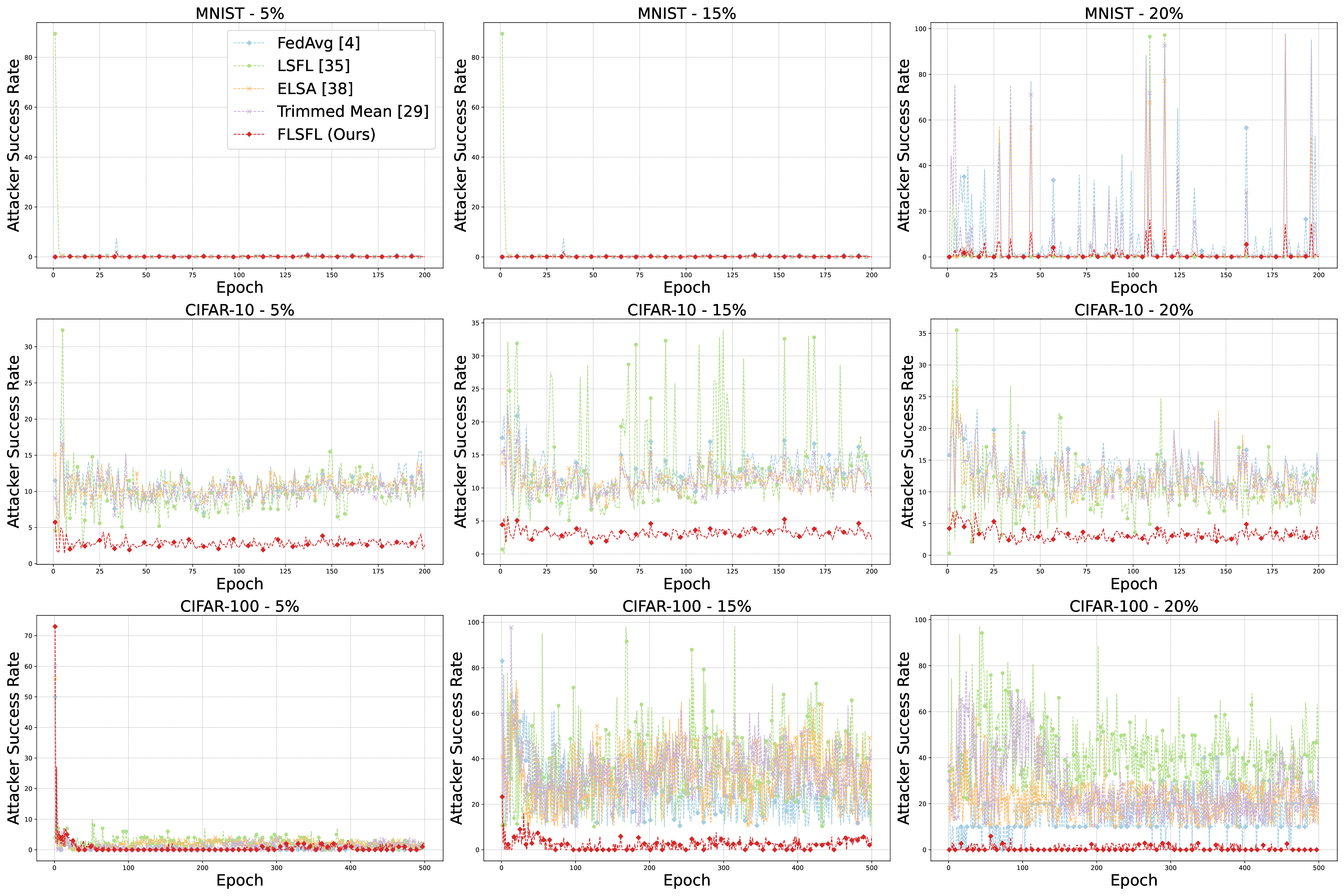}
    \caption{Attacker success rate over training epochs across different aggregation protocols under label-flipping attacks. The success rate reflects the proportion of malicious updates that successfully influence the global model. DSFL demonstrates a rapid suppression of adversarial influence, reaching near-zero success rates within the first few epochs and maintaining robustness throughout training. In contrast, FedAvg remains highly vulnerable, and LSFL shows partial resistance with fluctuating success rates, particularly under non-IID data distributions.}
    \label{fig: attacker_success_rate}
\end{figure*}

To rigorously assess the effectiveness and robustness of the proposed DSFL scheme, we conducted a series of experiments under Byzantine settings across three standard image classification benchmarks MNIST, CIFAR-10, and CIFAR-100, using both IID and non-IID data distributions. Our evaluation focuses on test accuracy degradation under label-flipping attacks, comparing DSFL with widely adopted aggregation protocols including FedAvg, Trimmed Mean, Median, DPBFL, LSFL, and homomorphic encryption-based (HE) approaches. The results are presented in Table~\ref{Table: Performance Comparison} and Table~\ref{tab:test_accuracy_against_attackers}, with corresponding trends visualised in Figures~\ref{fig: DSFL_comparison_all} and~\ref{fig: attacker_success_rate}.

Table~\ref{Table: Performance Comparison} illustrates the accuracy of each method under a fixed 20\% Byzantine participant ratio. FedAvg, despite its simplicity and wide usage in FL, exhibits substantial vulnerability to malicious participants. On the MNIST dataset, FedAvg achieves 80.67\% accuracy under IID data, which degrades slightly to 78.75\% under a non-IID distribution. However, its performance collapses dramatically on more complex datasets. On CIFAR-100 (non-IID), FedAvg reaches only 28.10\% accuracy, a level insufficient for any meaningful classification task. This degradation corroborates earlier findings by Blanchard et al.\cite{blanchard2017machine} and Bagdasaryan et al.\cite{backdoor}, who demonstrated that FedAvg is particularly fragile in heterogeneous and adversarial settings due to its unweighted averaging of all client updates, including malicious ones.

Secure aggregation protocols such as SecAgg~\cite{aggregation2017} and homomorphic encryption-based approaches~\cite{He2018} offer some improvement in preserving accuracy under attack but still fall short in highly non-IID contexts. Although HE-based methods achieve a moderate gain in robustness (e.g., 58.17\% on CIFAR-100 non-IID), their excessive computational and communication overheads make them impractical for real-time or resource-constrained deployments. Robust statistical methods like Trimmed Mean and Median perform better, particularly in datasets like CIFAR-10 and CIFAR-100. However, these methods rely on hard-coded trimming thresholds or majority assumptions that may fail under dynamic or adaptive attack patterns. Notably, Trimmed Mean achieves 60.87\% accuracy on CIFAR-10 non-IID, but only 50.43\% on CIFAR-100 non-IID, revealing its limited generalizability.

DSFL consistently outperforms all baselines across datasets and data partitions. On MNIST (IID), DSFL achieves 96.12\% accuracy despite the presence of 20\% Byzantine participants. On CIFAR-10 and CIFAR-100, DSFL maintains over 95\% and 68\% accuracy, respectively, with marginal drops in non-IID conditions. Compared to LSFL~\cite{LSFL}, which was previously one of the strongest robust FL baselines, DSFL provides comparable or slightly better performance with improved stability, particularly on CIFAR-100, where it surpasses LSFL by over 1.5\% under a non-IID distribution. These results suggest that the synergy between local differential privacy, the dual-server secure aggregation design, and fairness-enhanced participant selection enables DSFL to simultaneously preserve model utility and suppress malicious influence.

Table~\ref{tab:test_accuracy_against_attackers} and Figure~\ref{fig: attacker_success_rate} further illustrate the resilience of DSFL across increasing levels of adversarial intensity. While FedAvg degrades sharply with just 10\% Byzantine clients—dropping to as low as 9.39\% on CIFAR-100 non-IID, DSFL sustains robust accuracy even at 30\% adversaries. On CIFAR-10, DSFL maintains 97.25\% under IID and 92.53\% under non-IID distribution with 30\% malicious participants, underscoring its high tolerance to adversarial drift. This ability to scale robustly with adversarial pressure stands in contrast to most prior defences, including DPBFL~\cite{DPBFL} and ELSA~\cite{elsa}, which rely either on strong assumptions of client behaviour or complex cryptographic operations that are computationally demanding.

Figure~\ref{fig: DSFL_comparison_all} complements these findings by visualising the attack-specific test accuracy across 5\%, 10\%, and 20\% Byzantine client ratios. The curves reveal that DSFL maintains superior performance across all attack levels and datasets. In MNIST and CIFAR-10, its performance is almost indistinguishable from the no-attack baseline for up to 10\% adversaries. Additionally, the attack success rate—measured through the attack impact on clean data—is shown to remain negligible throughout training in both IID and non-IID conditions. Notably, in non-IID settings, where many defence schemes lose their effectiveness due to high gradient variance, DSFL stabilises quickly and suppresses malicious gradients through noise injection and masked aggregation, thereby neutralising their impact before convergence.

Figure~\ref{fig: attacker_success_rate} illustrates the attacker success rate over training epochs across various aggregation protocols under adversarial conditions. This metric quantifies the proportion of poisoned inputs (or label-flipped samples) that successfully influence the global model, thereby serving as a direct measure of the system’s vulnerability to Byzantine behaviour.

As observed in the figure, FedAvg exhibits a consistently high attacker success rate throughout the training process, particularly in non-IID scenarios. This is expected, as FedAvg performs naïve averaging without any robustness enhancement, making it highly susceptible to gradient manipulation. The attack success rate remains above 60\% across all rounds, confirming the attacker’s persistent influence on the global model.

In contrast, LSFL introduces filtering mechanisms based on update similarity and exhibits moderate resistance to adversarial behaviour. Its attacker success rate declines after several rounds of training but still fluctuates due to its sensitivity to gradient divergence in non-IID settings.

The proposed DSFL method demonstrates the most effective suppression of attacker success. Across all datasets and data distributions, its success rate drops to near zero within the first 5–10 communication rounds and remains negligible thereafter. This suggests that DSFL not only mitigates the impact of malicious clients early in training but also maintains stability and robustness throughout. This resilience can be attributed to the combined use of differential privacy noise, fair participant selection, and secure aggregation, which jointly obscure and filter malicious updates before they can affect the global model.

These results reinforce the findings in Table~\ref{Table: Performance Comparison} and Figure~\ref{fig: attacker_success_rate}, showing that DSFL does not merely preserve accuracy under attack but actively suppresses adversarial success, outperforming prior methods such as LSFL, DPBFL, and HE-based schemes both in robustness and efficiency.

Overall, these results demonstrate the capability of DSFL to provide strong Byzantine robustness while preserving high model accuracy across a range of FL scenarios. By avoiding the heavy computational burden of homomorphic encryption and the heuristic nature of statistical outlier rejection, DSFL offers a principled and scalable solution for secure and privacy-preserving FL. Its design ensures compatibility with non-IID data, which is prevalent in real-world FL deployments such as mobile and edge environments, making it a practical and deployable alternative to both classical and state-of-the-art defences. 

\subsection{Determination of Optimal Hyper-parameters}
To ensure fair comparison and optimal performance across all evaluated aggregation protocols, we conducted an extensive hyper-parameter search for key variables such as local learning rate, number of local epochs, batch size, and noise scale for differential privacy, where applicable. For DSFL, the learning rate was selected from the range {0.01, 0.001, 0.0005}, with 0.001 yielding the most stable convergence across datasets. The number of local epochs was fixed at 5 after observing diminishing returns and potential overfitting beyond this threshold. A batch size of 10 was used consistently to maintain training granularity, while the noise scale for differential privacy was tuned to balance privacy protection and model utility.

All baseline methods were fine-tuned within their recommended settings as reported in their original papers to avoid bias. For example, in DP-based methods such as DPBFL, we adopted the noise multiplier and clipping norm suggested in~\cite{DPBFL}. For HE-based methods, cryptographic parameters such as key size and ciphertext modulus were selected based on the configurations outlined in~\cite{He2018} to ensure feasible runtime. This careful calibration ensured that each method was evaluated under its best-performing configuration, thereby enabling a fair and rigorous comparison.

\subsection{Efficiency and Overhead Analysis}

\begin{figure}[t]
    \centering
    \includegraphics[width=1\linewidth]{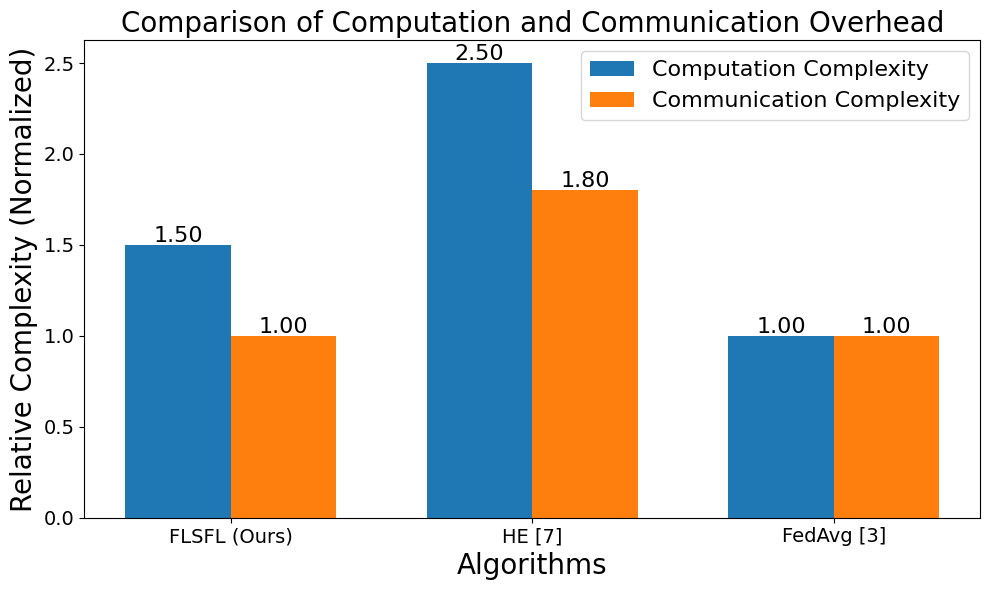}
    \caption{Comparison of computational and communication complexities for DSFL, FedAvg, and HE-based solutions~\cite{Wei2020}. DSFL achieves strong defence with a marginal increase in complexity compared to FedAvg and substantially lower overhead than HE-based protocols.}
    \label{fig:complexity}
\end{figure}

We evaluate the computational and communication efficiency of DSFL in comparison with two existing secure aggregation baselines: FedAvg~\cite{McMahan2016} and a homomorphic encryption (HE)-based scheme~\cite{Wei2020}. The comparison is summarised in Table~\ref{runtime-table} and visualised in Figure~\ref{fig:complexity}.

FedAvg~\cite{McMahan2016} is known for its low overhead. Its computational complexity includes local model training and optional encryption:
\[
\mathcal{O}(T_{\text{tr}}) + \mathcal{O}(mT_{\text{enc}})
\]
where \( T_{\text{tr}} \) denotes the local training time, \( T_{\text{enc}} \) is the per-client encryption cost, and \( m \) is the number of participants. The corresponding \textit{communication complexity} is:
\[
\mathcal{O}(m|w_{\text{enc}}|)
\]
with \( |w_{\text{enc}}| \) denoting the size of the encrypted model.

In contrast, the proposed DSFL framework adopts a dual-server secure aggregation protocol using additive noise. Its \textit{computational complexity} is:
\[
\mathcal{O}(T_{\text{tr}}) + \mathcal{O}(mT_{\text{ns}}) + \mathcal{O}(2mT_{\text{add}})
\]
where \( T_{\text{ns}} \) is the time for local noise generation, and \( T_{\text{add}} \) denotes the time for secure modular addition. DSFL avoids heavy cryptographic operations, enabling lightweight execution.

The \textit{communication overhead} is:
\[
\mathcal{O}(2m|w|)
\]
as each client transmits masked model updates to two servers. Although this doubles the number of transmissions compared to FedAvg, the use of unencrypted (masked) values significantly reduces the payload size compared to HE-based approaches.

Figure~\ref{fig:complexity} illustrates that DSFL achieves competitive runtime and communication cost—only slightly higher than FedAvg but considerably lower than HE-based methods. Empirical results in Table~\ref{runtime-table} (for 10 clients) show that DSFL's per-round runtime is 55.9 ms with a communication payload of 1088 KB, demonstrating near-parity with FedAvg and significant gains over LSFL.

\begin{table}[]
    \caption{Runtime and Communication Overhead for 10 Clients on the MNIST Dataset}
    \label{runtime-table}
    \centering
    \small
    \begin{tabular}{lccc}
        \toprule
        \textbf{Method} & \textbf{Time/Round} & \textbf{Msg Size} & \textbf{CPU Load} \\
        \midrule
        FedAvg~\cite{McMahan2016} & 22.4 ms & 512 KB & Low \\
        HE-Based~\cite{Wei2020} & 1020.3 ms & 13.2 MB & High \\
        LSFL~\cite{LSFL} & 58.1 ms & 1100 KB & Medium \\
        DSFL (Ours) & \textbf{55.9 ms} & \textbf{1088 KB} & Medium \\
        \bottomrule
    \end{tabular}
\end{table}

In summary, DSFL strikes an effective balance between security and system efficiency. It offers significantly improved robustness against adversarial threats, with low runtime and communication overhead, making it practical for real-world, privacy-sensitive FL deployments.

\subsection{Security Analysis}

We analyse the security guarantees of the proposed DSFL scheme under the assumptions that the Service Provider (SP) and the Trusted Authority (TP) are honest-but-curious and do not collude. Additionally, we consider the presence of Byzantine Device participants capable of launching adversarial attacks.

\textbf{Theorem 1:} \textit{Neither the SP, TP, nor Byzantine $P$s can infer the sensitive local model updates of honest-but-curious $P$s.}

\textit{Proof Sketch:} In DSFL, each $P$ splits its local model update into two random shares and sends them to the SP and TP, respectively. During the secure aggregation process, SP may send CPG TP to support Byzantine robustness. However, since SP and TP receive disjoint and partial information, and are non-colluding, they cannot reconstruct the full model update $\textbf{w}_i$ of any participant. Similarly, each $P$ is aware only of its update and the global model, and cannot infer updates of others. Therefore, DSFL preserves model privacy against both external and internal adversaries.

\textbf{Theorem 2:} \textit{The dual-server Secure Model Update protocol is secure as long as SP and TP do not collude.}

\textit{Proof Sketch:} From SP’s perspective, its view includes $(\textbf{w}_i^{(2)}, \bar{\textbf{w}}, CPG_{i,j}, PMC_{i,j}, \textbf{w})$, where $CPG_{i,j}$ non-comprizable and $\textbf{w} = \textbf{w}^{(1)} + \textbf{w}^{(2)}$ is the intermediate aggregated global model. Given access only to $CPG_{i,j}$ and $\textbf{w}$, SP cannot infer $\textbf{w}^{(1)}$ or reconstruct any individual $w_i$. Likewise, the view of TP includes $(\textbf{w}_i^{(2)},  PMC_{i,j}, )$. The views of both SP and TP are therefore computationally indistinguishable from simulated versions, implying no privacy leakage under the semi-honest model.

\textbf{Theorem 3:} \textit{DSFL is robust against Byzantine $P$s as long as the majority of $P$s are honest-but-curious.}

\textit{Proof Sketch:} DSFL employs a Euclidean distance-based filtering mechanism to select the $k$ most reliable local model updates. Each $P$’s update is compared to the aggregated baseline using $d_{G_i} =  \parallel \bar{\mathbf{w}} - \bar{\mathbf{w}}_{G_i} \parallel ^2_2.$, and updates with the smallest deviation from the median are selected. This makes it difficult for Byzantine $P$s to inject poisoned updates without mimicking benign distributions, which in practice is statistically challenging. Given that the majority of $P$s are honest, malicious updates are unlikely to pass the selection filter. As a result, the scheme effectively mitigates poisoning and label-flipping attacks.

In summary, DSFL achieves privacy-preserving and Byzantine-robust aggregation without relying on heavy cryptographic primitives or trusted hardware. Its dual-server design prevents inference attacks by curious servers and limits the influence of malicious participants through robust filtering.

\subsection{Convergence Analysis}

In this section, we rigorously analyse the convergence properties of the proposed DSFL framework. Unlike traditional single-server FL systems, DSFL employs a dual-server architecture (Trust Provider and Service Provider) with secret sharing and participant filtering, which introduces unique structural advantages. We show that despite these privacy-preserving and adversarial-robust features, DSFL maintains the optimal convergence rate of $\mathcal{O}(1/T)$ even under non-IID and partially Byzantine settings.

\subsubsection{Preliminaries and Problem Setup}

Let the global optimisation objective be defined as:
\begin{equation}
    \min_{\mathbf{w}} F(\mathbf{w}) = \frac{1}{N} \sum_{p=1}^{N} F_p(\mathbf{w}),
\end{equation}
where $F_p(\mathbf{w}) = \mathbb{E}_{\xi_p}[\ell(\mathbf{w}; \xi_p)]$ is the local empirical risk over participant $P$’s data sampled via $\xi_p \sim D_p$. The local SGD update per epoch is:
\begin{equation}
    \mathbf{w}_p \leftarrow \mathbf{w}_{p} - \eta \nabla \ell(\mathbf{w}_{p}; \xi_p).
\end{equation}

\subsubsection{Assumptions}

We adopt the following standard assumptions for convergence~\cite{SCAFFOLD2019, yin2018byzantinerobust}, \cite{Stich2018LocalSGD}:

\begin{assumption}[Smoothness]
Each local loss $F_p$ is $L$-smooth:
\[
\|\nabla F_p(\mathbf{w}) - \nabla F_p(\mathbf{v})\| \leq L \|\mathbf{w} - \mathbf{v}\|, \quad \forall \mathbf{w}, \mathbf{v}.
\]
\end{assumption}

\begin{assumption}[Strong Convexity]
Each $F_p$ is $\mu$-strongly convex:
\[
(\nabla F_p(\mathbf{w}) - \nabla F_p(\mathbf{v}))^T (\mathbf{w} - \mathbf{v}) \geq \mu \|\mathbf{w} - \mathbf{v}\|^2.
\]
\end{assumption}

\begin{assumption}[Bounded Variance]
The stochastic gradient has bounded variance:
\[
\mathbb{E}_{\xi_p} \| \nabla \ell(\mathbf{w}; \xi_p) - \nabla F_p(\mathbf{w}) \|^2 \leq \sigma^2.
\]
\end{assumption}

\subsubsection{Secure dual-Server Aggregation Protocol}

Each client $P_i$ splits their local model update $\mathbf{w}_p$ into two shares:
\[
\mathbf{w}_p^{(1)} = \mathbf{w}_p + \zeta_p, \quad
\mathbf{w}_p^{(2)} = \mathbf{w}_p - \zeta_p,
\]
where $\zeta_p \sim \mathcal{N}(0, g^2)$ is random noise known only to the client. TP and SP compute:
\[
z_1 = \frac{1}{N} \sum_{i=1}^{N} \mathbf{w}_p^{(1)}, \quad
z_2 = \frac{1}{N} \sum_{i=1}^{N} \mathbf{w}_p^{(2)},
\]
and the global model is reconstructed securely by SP as:
\[
\bar{\mathbf{w}} = \frac{z_1 + z_2}{2} = \frac{1}{N} \sum_{i=1}^{N} \mathbf{w}_p.
\]

\subsubsection{Main Convergence Result}

Let $\mathbf{w}^{(t)}$ be the global model at round $t$ and $E$ the number of local steps. Then under a learning rate schedule $\eta_t = \frac{\alpha}{t + \gamma}$, the model update satisfies:
\begin{align}
\mathbb{E} \left\| \mathbf{w}^{(t+1)} - \mathbf{w}^* \right\|^2
&\leq \left(1 - \frac{\alpha \mu}{t + \gamma} \right) \mathbb{E} \left\| \mathbf{w}^{(t)} - \mathbf{w}^* \right\|^2 \notag \\
&\quad + \frac{\alpha^2 (\sigma^2 E / K + E^2 G^2)}{\mu^2 (t + \gamma)^2},
\end{align}
where $G$ bounds the magnitude of adversarial updates and $K$ is the number of honest participants selected after filtering. This leads to a convergence rate:
\[
\mathbb{E}[F(\mathbf{w}^{(T)})] - F^* \leq \mathcal{O}\left(\frac{1}{T}\right).
\]

\subsubsection{Byzantine Filtering and Impact}

DSFL applies a deviation-based filtering mechanism, selecting the top $K' = K - \beta K$ participants with minimum deviation from the median update:
\[
\tilde{\mathbf{w}}^{(t+1)} = \frac{1}{K'} \sum_{i \in S} \mathbf{w}_p^{(t+1)}, \quad |S| = K'.
\]
This ensures adversarial resilience without sacrificing convergence guarantees:
\[
\mathbb{E}[F(\tilde{\mathbf{w}}^{(T)})] - F^* = \mathcal{O}\left(\frac{1}{T}\right).
\]

\subsubsection{Conclusion}

DSFL achieves the optimal convergence rate for federated optimization under convexity, while providing robustness against adversarial behavior and preserving update privacy via secure dual-server aggregation. This theoretical foundation supports its practical deployment in distributed, edge-device scenarios.


\section{Conclusion}
\label{Section: Conclusion}

This paper presented DSFL, a Lightweight, Dual-Server Byzantine-Resilient FL Framework that simultaneously addresses three critical challenges in FL: privacy leakage, Byzantine robustness, and communication overhead. By leveraging a novel dual-server architecture with non-colluding Trust and Service Providers, DSFL securely aggregates model updates without relying on costly cryptographic primitives such as Homomorphic Encryption. Its group-based deviation filtering and dynamic credit mechanism allow robust participant selection and fair aggregation, even under strong adversarial presence. Unlike existing approaches, DSFL achieves strong privacy protection through additive noise sharing, high resilience to Byzantine participants via top-$k$ filtering, and computational efficiency suitable for resource-constrained edge deployments. Our security analysis shows that the protocol mitigates information leakage even under honest-but-curious servers. Extensive experiments on MNIST, CIFAR-10, and CIFAR-100, under both IID and non-IID settings with up to 30\% adversaries, demonstrate that DSFL consistently outperforms state-of-the-art baselines, including LSFL, HE-based schemes, and DPBFL. It preserves model utility with minimal overhead, maintaining accuracies above 90\% even under severe poisoning attacks, while reducing attacker success rates to near zero. Overall, DSFL bridges the gap between secure, robust, and efficient FL, making it a strong candidate for real-world deployment in edge and IoT scenarios. In future work, we plan to extend DSFL to support adaptive trust scoring, real-time participant churn, and scalability to large-scale heterogeneous FL systems. Our findings pave the way for scalable, trustworthy FL in critical domains such as healthcare, finance, and IoT, where both privacy and robustness are paramount.


\bibliographystyle{IEEEtran}
\bibliography{references}
\vspace{-2em}
\begin{IEEEbiography}[{\includegraphics[width=1in,height=1.25in,clip,keepaspectratio]{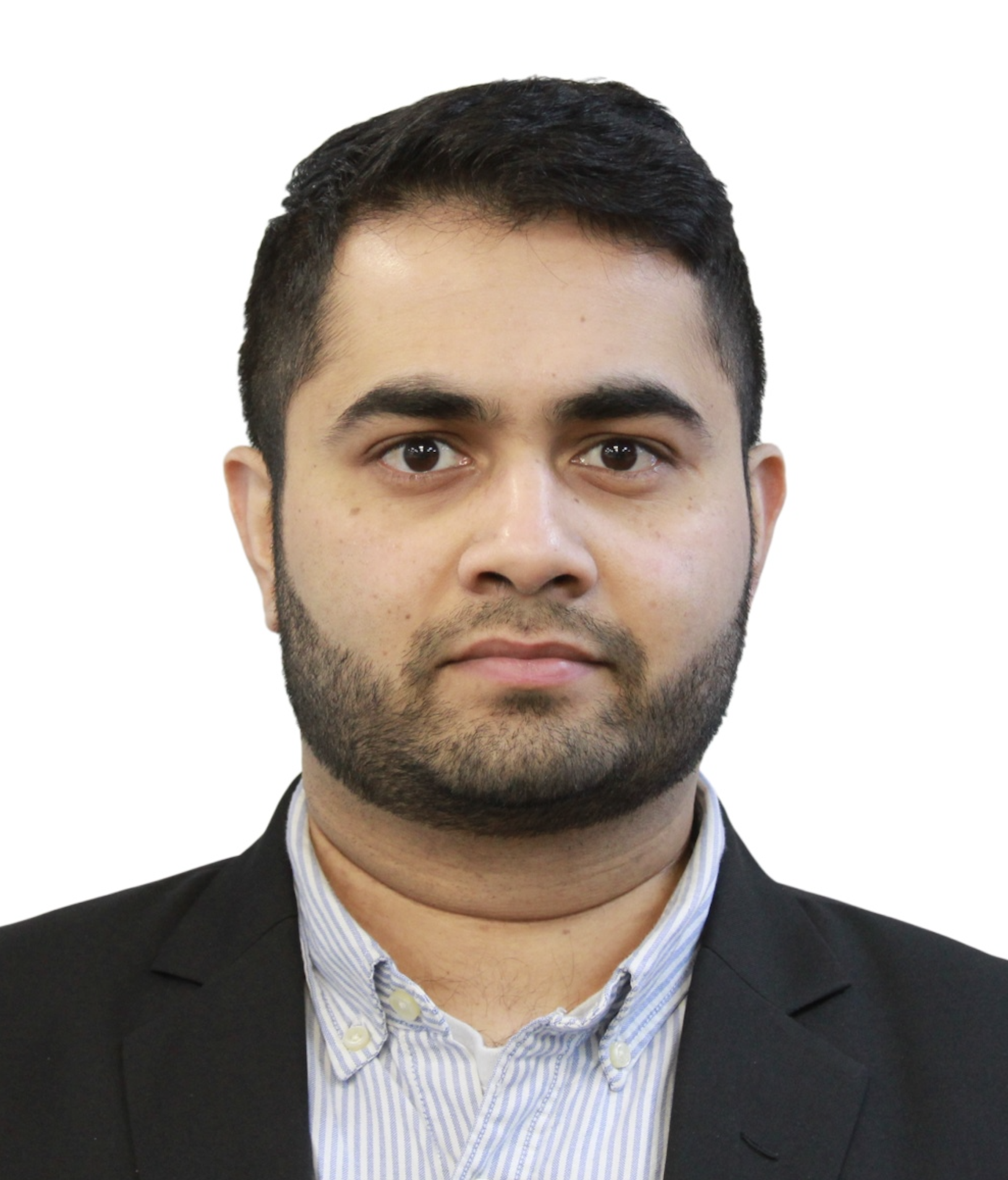}}]{Charuka Herath} is currently pursuing a PhD in AI and Cybersecurity at Loughborough University, London. He is an alumnus of the University of Moratuwa, Sri Lanka, and serves as a Research Assistant on a post-quantum cryptography project with Airbus Defence and Space, in collaboration with Cardiff Metropolitan University and Loughborough University. Furthermore, he works as an ML developer for a collaborative project under a grant on Blockchain, AI and Sanctions. Additionally, he is an Associate Lecturer in Computing at Arden University, London. Before this, he worked as a Senior Software Engineer at Sysco LABS and WSO2 Private Limited from 2020 to 2022. His research interests include wireless distributed learning, large language models, one-shot learning, and cybersecurity. 
\end{IEEEbiography}

\vspace{-1.5em}
\begin{IEEEbiography}[{\includegraphics[width=1in,height=1.25in,clip,keepaspectratio]{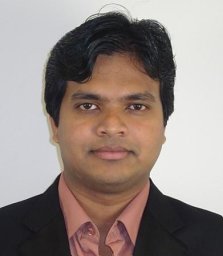}}]{Yogachandran Rahulamathavan} is a Reader in Cybersecurity and Privacy and joined Loughborough University, London in 2016. After obtaining a Ph.D. in Signal Processing from Loughborough University in 2012, Rahul joined the Information Security Group at City, University of London as a Research Fellow to lead signal processing in the encrypted domain research theme. During his time as a research fellow, he was a security and privacy work package leader for a Large-Scale Integrated Project, SpeechXrays, funded by the European Commission. Since April 2016, he joined Loughborough University's postgraduate campus in London as a lecturer, where he was promoted to senior lecturer in January 2020 and to Reader in January 2024. He is the module leader for Cybersecurity and Forensics, Principles of Artificial Intelligence and Data Analytics and Information Management modules. He is one of the recipients of the British Council's UK-India research funding in 2017 and successfully led a project between Lboro, City and IIT Kharagpur. Currently, Rahul leads a team of three Ph.D. students and serves as the principal investigator for an industry-funded project supported by Airbus Defence. Rahul is a Programme Director for MSc Cyber Security and Data Analytics at Loughborough University, London.
\end{IEEEbiography}

\vspace{-1.5em}
\begin{IEEEbiography}[{\includegraphics[width=1in,height=1.25in,clip,keepaspectratio]{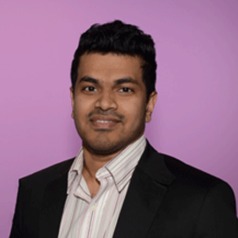}}]{Varuna De Silva} is a Professor Machine Intelligence at Loughborough University London. He received his PhD in Electronic Engineering from the University of Surrey in 2011. He developed award-winning algorithms at Apical Ltd, including patented CMOS camera technology now used in hundreds of millions of devices worldwide. Since 2016, he has been at Loughborough University, where he is a Professor of Machine Intelligence. His research focuses on neuromorphic AI, multi-agent reinforcement learning, multimodal computer vision, and simulation for real-world engineering. Professor De Silva is Associate Editor of IEEE Access, a Fellow of the Higher Education Academy, a member of BCS and IET, and Thematic Lead on AI and Digital Technologies for the UK Parliament.
\end{IEEEbiography}

\vspace{-1.5em}
\begin{IEEEbiography}[{\includegraphics[width=1in,height=1.25in,clip,keepaspectratio]{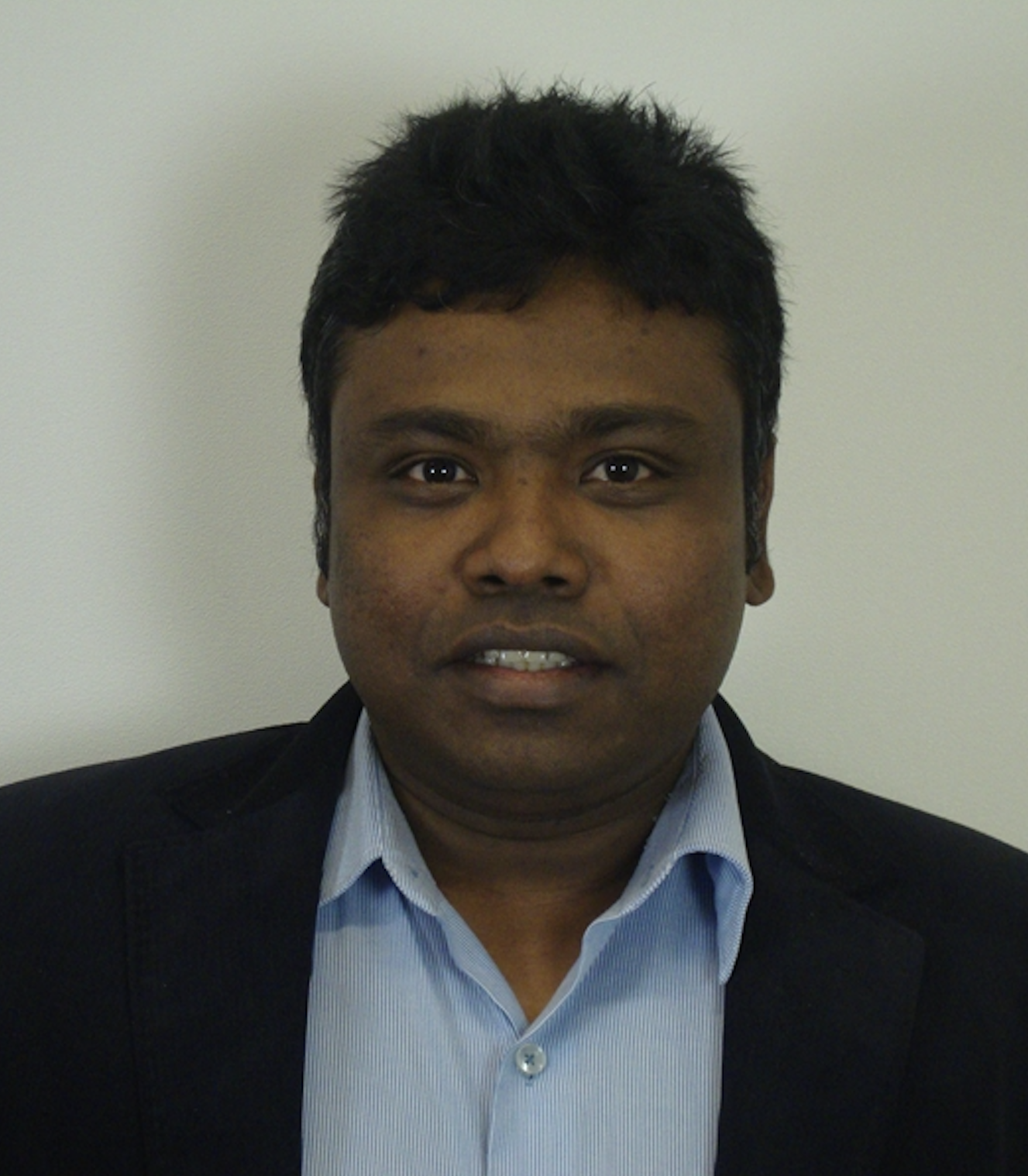}}]{Sangarapillai Lambotharan} is Professor of Signal Processing and Communications and the Director of the Institute for Digital Technologies at Loughborough University, London. He received his PhD in signal processing from Imperial College London, U.K., in 1997. He was a Visiting Scientist with the Engineering and Theory Centre, Cornell University, USA, in 1996. Until 1999, he was a Post-Doctoral Research Associate at Imperial College London. From 1999 to 2002, he was with the Motorola Applied Research Group, U.K., where he investigated various projects, including physical link layer modelling and performance characterization of 2.5G and 3G networks. He was with King’s College London and Cardiff University as a Lecturer and a Senior Lecturer, respectively, from 2002 to 2007. His current research interests include 6G networks, MIMO, blockchain, machine learning, and network security. He has authored more than 275 journal articles and conference papers in these areas. He is a Fellow of IET and a Senior Member of IEEE. He serves as a Senior Area Editor for IEEE Transactions on Signal Processing.
\end{IEEEbiography}

\end{document}